\documentclass[11pt,a4paper]{article}
\usepackage{mciteplus}
\usepackage[numbers, square, comma, sort&compress]{natbib}
\usepackage{amssymb}
\usepackage{amsthm}
\usepackage{amstext}
\usepackage{amsmath}
\usepackage{amsfonts}
\usepackage{verbatim}
\usepackage{geometry}
\usepackage{latexsym}
\usepackage[dvips]{graphicx}
\usepackage{t1enc}
\usepackage{graphicx}
\usepackage[all]{xy}
\usepackage{makeidx}
\usepackage{slashed}
\usepackage{multicol}

\def\spa#1.#2{\left\langle#1\,#2\right\rangle}
\def\spb#1.#2{\left[#1\,#2\right]}
\def\spash#1.#2{\spa{\smash{#1}}.{\smash{#2}}}
\def\spbsh#1.#2{\spb{\smash{#1}}.{\smash{#2}}}
\def\sand#1.#2.#3{%
\left\langle\smash{#1^{-}}{\vphantom1}\right|{#2}%
\left|\smash{#3^{-}}{\vphantom1}\right\rangle}
\def\sandp#1.#2.#3{%
\left\langle\smash{#1^{-}}{\vphantom1}\right|{#2}%
\left|\smash{#3^{+}}{\vphantom1}\right\rangle}
\def\sandpp#1.#2.#3{%
\left\langle\smash{#1^{+}}{\vphantom1}\right|{#2}%
\left|\smash{#3^{+}}{\vphantom1}\right\rangle}
\def\sandpm#1.#2.#3{%
\left\langle\smash{#1^{+}}{\vphantom1}\right|{#2}%
\left|\smash{#3^{-}}{\vphantom1}\right\rangle}
\def\sandmp#1.#2.#3{%
   \left\langle\smash{#1^{-}}{\vphantom1}\right|{#2}%
    \left|\smash{#3^{+}}{\vphantom1}\right\rangle}

\def\ketm#1{|\smash{#1}\vphantom{1}]}
\def\ketp#1{|\smash{#1}\vphantom{1}\rangle}
\def\tlambdap{{\tilde\lambda'}{}}

\def\ssand#1.#2.#3{%
\left\langle\smash{#1}{\vphantom1}\right|{#2}%
\left|\smash{#3}{\vphantom1}\right]}
\def\ssandp#1.#2.#3{%
\left\langle\smash{#1}{\vphantom1}\right|{#2}%
\left|\smash{#3}{\vphantom1}\right\rangle}
\def\ssandpp#1.#2.#3{%
\left\langle\smash{#1}{\vphantom1}\right|{#2}%
\left|\smash{#3}{\vphantom1}\right\rangle}

\def\proj{\flat}
\def\projdot#1.#2{k_{#1}^\proj\cdot k_{#2}^\proj}
\def\sandff#1.#2.#3{%
\left\langle\smash{#1^{\proj,-}}{\vphantom1}\right|{#2}%
\left|\smash{#3^{\proj,-}}{\vphantom1}\right\rangle}
\def\sandnf#1.#2.#3{%
\left\langle\smash{#1^{-}}{\vphantom1}\right|{#2}%
\left|\smash{#3^{\proj,-}}{\vphantom1}\right\rangle}
\def\sandfn#1.#2.#3{%
\left\langle\smash{#1^{\proj,-}}{\vphantom1}\right|{#2}%
\left|\smash{#3^{-}}{\vphantom1}\right\rangle}

\def\ketfm#1{|\smash{#1}\vphantom{1}^{\proj}]}
\def\ketfp#1{|\smash{#1}\vphantom{1}^{\proj}\rangle}

\def\eqn#1{eq.~(\ref{#1})}
\def\Eqn#1{Eq.~(\ref{#1})}
\def\eqns#1#2{eqs.~(\ref{#1}) and~(\ref{#2})}

\def\tlambdap{{\tilde\lambda'}{}}
\def\tree{{\rm tree}}

\def\onehalf{\frac12}
\def\Sol{{\cal S}}
\def\xib{\bar\xi}
\def\nn{\nonumber}
\def\Global{{\cal G}}
\def\ji{J_{\oint_{}^{}}}
\def\spa#1.#2{\left\langle#1\,#2\right\rangle}
\def\spb#1.#2{\left[#1\,#2\right]}
\def\Ord{{\cal O}}

\newif\ifdraft
\draftfalse
\newif\ifpreprint
\preprinttrue

\numberwithin{equation}{section}

\begin{document}
\begin{titlepage}
\hbox{CERN-PH-TH/2013-191}  \hbox{NIKHEF/2013--026}
\hbox{Saclay IPhT--T13/204}
\vskip 30mm

\begin{center}
\Large{\textbf{Maximal Unitarity for the Four-Mass Double Box}}
\end{center}

\vskip 6mm

\begin{center}
Henrik Johansson$^a$, David A. Kosower$^b$ and Kasper J. Larsen$^{c,d}$\\[6mm]

\textit{$^a$Theory Group, Physics Department, CERN, CH--1211 Geneva 23, Switzerland}\\[1mm]
\textit{$^b$Institut de Physique Th\'eorique, CEA-Saclay, F--91191 Gif-sur-Yvette cedex, France}\\[1mm]
\textit{$^c$Nikhef, Theory Group, Science Park 105, NL--1098 XG Amsterdam, The Netherlands}\\[1mm]
\textit{$^d$School of Natural Sciences, Institute for Advanced Study, Princeton, NJ 08540, USA}\\[4mm]

{\texttt{Henrik.Johansson@cern.ch, David.Kosower@cea.fr, Kasper.Larsen@nikhef.nl}}
\\[15mm]

\end{center}

\vskip -1mm

\begin{abstract}
\noindent We extend the maximal-unitarity formalism at two loops to
double-box integrals with four massive external legs.  These are
relevant for higher-point processes, as well as for heavy vector
rescattering, $VV\to VV$. In this formalism, the two-loop amplitude is
expanded over a basis of integrals. We obtain formulas for the
coefficients of the double-box integrals, expressing them as products
of tree-level amplitudes integrated over specific complex
multidimensional contours. The contours are subject to the consistency
condition that integrals over them annihilate any integrand whose
integral over real Minkowski space vanishes.  These include integrals
over parity-odd integrands and total derivatives arising from
integration-by-parts (IBP) identities.  We find that, unlike the zero-
through three-mass cases, the IBP identities impose no constraints on
the contours in the four-mass case.  We also discuss the algebraic
varieties connected with various double-box integrals, and show how
discrete symmetries of these varieties largely determine the constraints.
\end{abstract}

\end{titlepage}

\section{Introduction}

Last year's discovery~\cite{AtlasHiggs,CMSHiggs} by the ATLAS and CMS
collaborations of a Higgs-like boson completes the particle content of the
Standard Model.  Coupled with the absence to date of direct signals of
physics beyond the Standard Model, the discovery points towards an
important role for precision measurements in determining the scale of new
physics beyond the Standard Model.

Theoretical calculations at the LHC, whether for signals or backgrounds,
begin with the tree-level amplitudes required for leading-order (LO)
calculations in perturbative quantum chromodynamics (QCD).  Because the
strong coupling $\alpha_s$ is relatively large and runs quickly, LO
predictions suffer from strong dependence on the unphysical renormalization
and factorization scales and are thus not quantitatively reliable.
Next-to-leading order (NLO) is the lowest order in perturbation theory
which offers quantitatively reliable predictions.  These calculations
require one-loop amplitudes in addition to tree-level amplitudes with
higher multiplicity.  Recent years have seen major advances in NLO
calculations, especially for processes with several jets in the final
state~\cite{EMZW3j,BlackHatW3j,BlackHatZ3j,BlackHatW4j,BlackHatZ4j,Badger:2012pf,BlackHatW5j}.
 While the uncertainty left by scale variation
cannot be quantified in the same fashion as statistical uncertainties,
experience shows that it is of $\Ord$(10--15\%).

As combined experimental uncertainties in future measurements push below
this level, a comparison with theoretical calculations will require pushing
on to next-to-next-to-leading order (NNLO) accuracy.  Such studies will
require computation of two-loop amplitudes.  These computations form the
next frontier of precision QCD calculations.  The only existing
fully-exclusive NNLO jet calculations to date are for three-jet production
in electron--positron annihilation~\cite{NNLOThreeJet}.  These calculations
have been used to determine $\alpha_s$ to 1\% accuracy from jet data at
LEP~\cite{ThreeJetAlphaS}.  This extraction is competitive with other
determinations.  Beyond their use in seeking deviations in precision
experimental data from Standard-Model predictions, NNLO calculations will
also be useful at the LHC for improving our understanding of scale
stability in multi-scale processes such as $W$+multi-jet production, as
well as for providing honest theoretical uncertainty estimates for NLO
calculations.

The unitarity
method~\cite{UnitarityMethod,Bern:1995db,Zqqgg,DdimensionalI,BCFUnitarity,OtherUnitarity,Bootstrap,BCFCutConstructible,BMST,OPP,OnShellReview,Forde,Badger,DdimensionalII,BFMassive,BergerFordeReview,Bern:2010qa,Elvang:2013cua}
has made many previously-inaccessible one-loop calculations feasible.
Of particular note are processes with many partons in the final state.
The most recent development, applying generalized unitarity, allows
the method to be applied either analytically or purely
numerically~\cite{EGK,BlackHatI,CutTools,MOPP,Rocket,BlackHatII,CutToolsHelac,Samurai,WPlus4,NGluon,MadLoop}.
The numerical formalisms underlie recent software libraries and
programs that have been applied to LHC phenomenology.  In this
approach, the one-loop amplitude in QCD is written as a sum over a set
of basis integrals, with coefficients that are rational in external
spinors,
\begin{equation}
{\rm Amplitude} = \sum_{j\in {\rm Basis}}
  {\rm coefficient}_j \times {\rm Integral}_j +
{\rm Rational}\,.
\label{BasicEquation}
\end{equation}
The integral basis for amplitudes with massless internal lines
contains box, triangle, and bubble integrals in addition to purely
rational terms (dropping all terms of $\mathcal{O}(\epsilon)$ in the
dimensional regulator).  The coefficients are calculated from products
of tree amplitudes, typically by performing contour integrals via
discrete Fourier projection.  In the Ossola--Papadopoulos--Pittau (OPP)
approach~\cite{OPP}, this
decomposition is carried out at the integrand level rather than at
the level of integrated expressions.

Higher-loop amplitudes can also be written in the form given in
\eqn{BasicEquation}.  As at one loop, one can carry out such a
decomposition at the level of the integrand.  This generalization of the
OPP approach has been pursued by Mastrolia and
Ossola~\cite{MastroliaOssola} and collaborators, and also by Badger,
Frellesvig, and~Zhang~\cite{BadgerFZ}.  The reader should consult
refs.~\cite{Zhang:2012ce,Mastrolia:2012an,Kleiss:2012yv,Mastrolia:2012wf,Mastrolia:2012du,Huang:2013kh,Mastrolia:2013kca}
for further developments within this approach.  Arkani-Hamed and
collaborators have developed an integrand-level
approach~\cite{ArkaniHamed:2009dn,ArkaniHamed:2009dg,ArkaniHamed:2010kv,ArkaniHamed:2010gg,ArkaniHamed:2010gh,ArkaniHamed:2012nw}
specialized to planar contributions to the ${\cal N}=4$ supersymmetric
theory, but to all loop orders.

Within the unitarity method applied at the level of integrated expressions,
one can distinguish two basic approaches.  In a `minimal' application
of generalized unitarity, used in a number of prior applications
and currently pursued by Feng and Huang~\cite{Feng:2012bm}, one
cuts just enough propagators to break apart a higher-loop amplitude
into a product of disconnected tree amplitudes.
Each cut is then a product of tree amplitudes, but because not
all possible propagators are cut, each generalized cut will
correspond to several integrals, and
algebra will be required to isolate specific integrals and
their coefficients.
This approach does not require a
predetermined general basis of integrals; it can be determined
in the course of a specific calculation.
A number of calculations have been done this way, primarily
in the ${\cal N}=4$ supersymmetric gauge theory~\cite{Bern:1997nh,ABDK,
Bern:2005iz,Bern:2006vw,Bern:2006ew,Bern:2008ap,Kosower:2010yk,Bern:2011rj}, but including
several four-point calculations in QCD
and supersymmetric
theories with less-than-maximal supersymmetry~\cite{Bern:2000dn,Bern:2002tk,BernDeFreitasDixonTwoPhoton,Bern:2002zk,Bern:2003ck,TwoLoopSplitting,DeFreitas:2004tk}.
Furthermore, a number of recent multi-loop calculations in maximally supersymmetric gauge and gravity theories have used maximal cuts~\cite{Bern:2007ct, Bern:2008pv,LeadingSingularity,Bern:2010tq,Carrasco:2011hw,Carrasco:2011mn,Bern:2012uc}, without complete
localization of integrands.

\def\GenDisc{\mathop{\rm GenDisc}\nolimits}
We will use a more intensive form or `maximal'
form of generalized unitarity.  In this approach,
one cuts as many propagators as
possible, and further seeks to fully localize integrands onto global poles
to the extent possible.
In principle, this allows one to isolate individual integrals
on the right-hand side of the higher-loop analog of eq.~(\ref{BasicEquation}).
In previous papers~\cite{MaximalTwoLoopUnitarity,ExternalMasses},
we showed how to extract
the coefficients of double-box master integrals using
multidimensional contours
around global poles.
In this paper, we recast this operation as applying
{\it generalized discontinuity operators\/} (GDOs).
Each GDO corresponds to integrating the integrand of an amplitude or
an integral along a specified linear combination of
multidimensional contours
around global poles.
The GDOs generalize the operation of cutting via the Cutkosky-rule
replacement of propagators by on-shell delta functions.

Some of the contour integrations in a GDO put internal lines on shell,
equivalent to cutting propagators~\cite{Loops&Legs}.  This integration
will typically yield a Jacobian giving rise to poles in the
remaining degrees of freedom.  In the case of the double box, the
Jacobian allows one to fully localize the remaining degrees of freedom
through additional multidimensional contour integrals.  The integrand
is then fully localized at one of a set of global poles.  We call
these additional degrees of freedom `localization variables'.
We include
these additional dimensions of contours in the definition of the GDO.
This maximal-unitarity approach may be viewed as a generalization to
two loops of the work of Britto, Cachazo, and
Feng~\cite{BCFUnitarity}, and of Forde~\cite{Forde}.

The GDOs are constructed so that each one selects a specific master
integral,
\begin{equation}
\GenDisc_i\bigl({\rm Integral}_j\bigr) = \delta_{ij}\,.
\end{equation}
Applying it to \eqn{BasicEquation} then gives us an expression for
the corresponding coefficent,
\begin{equation}
{\rm coefficient}_j = \GenDisc_j\bigl({\rm Amplitude}\bigr)\,.
\end{equation}
The right-hand side will have the form of explicit contour integrals
over localization variables of a product of tree amplitudes; schematically,
\begin{equation}
{\rm coefficient}_j = \oint_{\Gamma_j} dz_{i} \;
\prod A^{\rm tree}(z_i)\,.
\end{equation}

The weights with which the contours $\Gamma_j$ surround the different global
poles are determined by a set of consistency equations.  These equations
require that integrals vanishing over the Minkowski
slice of complexified loop-momentum space are also annihiliated by
the GDOs, vanishing on the particular combinations of contours in
each and every GDO.

In this paper, we continue the maximal-unitarity approach of
refs.~\cite{MaximalTwoLoopUnitarity,ExternalMasses}, relying on the
global-pole analysis~\cite{Caron-HuotLarsen} of Caron-Huot and one of
the present authors.  At higher loops, the coefficients of the basis
integrals are no longer rational functions of the external spinors
alone, but will in general depend explicitly on the dimensional
regulator $\epsilon$.  We consider GDOs operating only on the
four-dimensional components of the loop momenta, and accordingly
extract only the leading terms, $\epsilon$-independent terms.  GDOs
operating on the full $D$-dimensional loop momenta would be required
to extract the remaining terms, and could presumably be used to obtain
the rational terms in \eqn{BasicEquation} as well.

At two loops and beyond, the number of master integrals for a given
topology will depend on the number and arrangement of external
masses~\cite{TwoLoopBasis}.  (See also recent work on a different
organization of higher-loop
integrals~\cite{Henn:2013pwa,Henn:2013tua,Henn:2013woa}.)  In previous
papers, we have considered double boxes with no external
masses~\cite{MaximalTwoLoopUnitarity} or with one, two, or three external
masses~\cite{ExternalMasses}.  In this article, we extend the GDO
construction to planar double boxes with four external masses.  We consider
both the general case with unequal masses, and one special case with pairs
of equal masses.  In the general case, there are four master integrals; in
the special case with an extra reflection symmetry, three.
S\o{}gaard~\cite{Sogaard} has constructed GDOs for the non-planar massless
double box.

As in previous work, we ensure the consistency of the GDOs by
requiring that they yield a vanishing result when applied to vanishing
integrals.  For the four-mass double box, it turns out that
non-trivial constraints arise only from parity-odd integrands;
integration-by-parts (IBP)
identities~\cite{IBP,Laporta,GehrmannRemiddi,LIdependent,AIR,FIRE,Reduze,SmirnovPetukhov}
give no additional constraints.  The symmetry requirement for the
special equal-mass case must also be imposed explicitly, and unlike
fewer-mass cases, does not emerge automatically from IBP equations.
We consider only two-loop master integrals with massless internal
lines.  We will not consider the generalization to massive internal
lines; but so long as there are sufficient massless internal lines to
have at least one chiral vertex, the integrand should still have
global poles, and we should expect the approach described here to
generalize smoothly.

This paper is organized as follows.  In sec.~\ref{sec:notation},
we present the parametrization of loop momenta we use for derivations.
In sec.~\ref{sec:four-mass_projectors}, we discuss the maximal-cut
equations for the four-mass double box along with the global poles,
and derive the GDOs for the four master integrals.  In
sec.~\ref{sec:Feynman_varieties}, we discuss constraint
equations and their symmetries for all double boxes from
an algebraic-geometry point of view.  We make some concluding
remarks in sec.~\ref{sec:Conclusions}.

\section{Loop-Momentum Parametrization}
\label{sec:notation}

\begin{figure}[t]
\begin{center}
\includegraphics[width=0.45\textwidth]{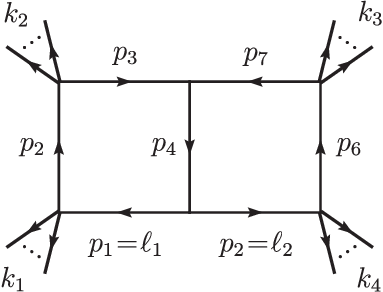}
\end{center}
{\vskip -6mm}
\caption{\small The double-box integral.}
\label{fig:DoubleBox}
\end{figure}

\def\ibar{{\bar\imath}}

We take over the same loop-momentum parametrization used in
ref.~\cite{ExternalMasses}.  This parametrization makes use of spinors
defined for massive external legs.  Such spinors correspond to massless
four-dimensional momenta, which we obtain using `mutually-projected'
kinematics.  This construction was previously used in the work of
OPP~\cite{OPP} and Forde~\cite{Forde} to extract
triangle and bubble coefficients at one loop.

For a given pair of external four-momenta
$(k_i,k_j)$, we require the mutually-projected momenta to satisfy
\begin{equation}
\begin{aligned}
k_{i}^{\proj,\mu} &= k_{i}^\mu -
\frac{k_{i}^2}{2 k_{i}\cdot k_{j}^\proj} k_{j}^{\proj,\mu}
  \,,\\
k_{j}^{\proj,\mu} &= k_{j}^\mu -
 \frac{k_{j}^2}{2 k_{j}\cdot k_{i}^\proj} k_{i}^{\proj,\mu}
  \,.
\end{aligned}
\label{MutualProjection}
\end{equation}
By construction, $k^\proj_{i}$ and $k^\proj_{j}$ are massless
momenta. Next, define
\begin{equation}
\rho_{ij} \equiv \frac{k_{i}^2}{2 k_{i}\cdot k_{j}^\proj}\,.
\end{equation}
We note that
\begin{equation}
k_{i}\cdot k_{j}^\proj
= k_{i}^\proj\cdot k_{j}
= k_{i}^\proj\cdot k_{j}^\proj\,;
\label{FlatRelations}
\end{equation}
and define
\begin{equation}
\gamma_{ij} \equiv 2 k_{i}^\proj\cdot k_{j}^\proj\,,
\end{equation}
so that $\rho_{ij}=k_i^2/\gamma_{ij}$. After using \eqn{MutualProjection}, we obtain a quadratic equation for $\gamma_{ij}$; its two solutions are
\begin{equation}
\gamma_{ij}^\pm =
  k_{i}\cdot k_{j} \pm
  \big[ (k_{i}\cdot k_{j})^2 - k_{i}^2 k_{j}^2\big]^{1/2}\,.
\end{equation}
If either momentum in the pair $(k_i,k_j)$ is massless, only one
solution survives.  \Eqn{FlatRelations} then gives us $\gamma_{ij} = 2
k_{i}\cdot k_{j}$\,.  Inverting \eqn{MutualProjection}, we obtain the
massless momenta
\begin{equation}
k_{i}^{\proj,\mu} = (1- \rho_{ij} \rho_{ji})^{-1}
   (k_{i}^{\mu} - \rho_{ij} \,k_{j}^{\mu})\,;
\end{equation}
swap $i \leftrightarrow j$ to obtain $k_{j}^{\proj,\mu}$.  In this
paper we work with two mutually-projected pairs: $(k_1,k_2)$ and
$(k_3,k_4)$.  This choice defines a set of `projected' massless momenta
$k_{i}^{\proj,\mu}$, $i=1,\ldots,4$, in terms of the external momenta,
$k_i$, and the sign choices in $\gamma_{12}^\pm$ and
$\gamma_{34}^\pm$.

With the projected momenta,
we adopt the following parametrization for the
double-box loop momenta as depicted in fig.~\ref{fig:DoubleBox},
\begin{equation}
\ell_1^\mu =\onehalf\ssand{\lambda_1}.{\gamma^\mu}.{\tlambdap_1}
+\zeta_1\eta_1^{\mu} \,,\qquad
\ell_2^\mu =\onehalf\ssand{\lambda_2}.{\gamma^\mu}.{\tlambdap_2}
+\zeta_2\eta_2^{\mu} \,,
\label{SpinorParametrization}
\end{equation}
where $\zeta_i$ are complex numbers, and the $\eta_i$ are null vectors
satisfying $\slashed{\eta}_1\ketp{\lambda_1} \neq 0\neq
\slashed{\eta}_1\ketm{\tlambdap_1}$
and $\slashed{\eta}_2\ketp{\lambda_2} \neq 0\neq
\slashed{\eta}_2\ketm{\tlambdap_2}$.  We introduce the $\zeta_i$
in order to compute Jacobian factors arising from the
change of variables in the double-box integral.  To obtain
on-shell momenta, we subsequently set $\zeta_i = 0$.

We write the
various loop spinors in \eqn{SpinorParametrization}
in terms of the spinors corresponding to
$(k_{1}^\proj,k_{2}^\proj)$ for $\ell_1$, and the spinors
corresponding to $(k_{3}^\proj,k_{4}^\proj)$ for $\ell_2$:
\begin{equation}
\begin{aligned}
\ketp{\lambda_1} &= \xi_1 \ketfp{1}
   + \xi_2 \frac{\spash{4^\proj}.{1^\proj}}
                {\spash{4^\proj}.{2^\proj}} \ketfp{2}\,,\\
\ketm{\tlambdap_1} &= \xi'_1 \ketfm{1}
   + \xi'_2 \frac{\spbsh{4^\proj}.{1^\proj}}
                 {\spbsh{4^\proj}.{2^\proj}} \ketfm{2}\,,\\
\ketp{\lambda_2} &= \xi_3 \frac{\spash{1^\proj}.{4^\proj}}
            {\spash{1^\proj}.{3^\proj}} \ketfp{3} + \xi_4 \ketfp{4}\,,\\
\ketm{\tlambdap_2} &= \xi'_3 \frac{\spbsh{1^\proj}.{4^\proj}}
            {\spbsh{1^\proj}.{3^\proj}} \ketfm{3} + \xi'_4 \ketfm{4}\,,
\end{aligned}
\label{BasicSpinorDefinitions}
\end{equation}
where the external spinors are defined via
$k_{i}^{\proj,\mu}=\ssand{i^\proj}.{\gamma^\mu}.{i^\proj}/2$.  Without
loss of generality, we can set two of the complex parameters to unity,
$\xi_1 = \xi_4 = 1$, as we will do throughout the paper.

Moreover, similarly to ref.~\cite{ExternalMasses}, we define the
following quantities:
\begin{equation}
\begin{alignedat}{2}
\xib'_1 \hspace{0.7mm}&\equiv\hspace{0.7mm}
  \frac{\gamma_{12} s_{12}-(\gamma_{12}+m_2^2) m_1^2}
                {\gamma_{12}^2-m_1^2 m_2^2} \,, \qquad
&\xib'_2 \hspace{0.7mm}&\equiv\hspace{0.7mm}
  -\frac{m_1^2 \bigl( s_{12}-\gamma_{12}-m_1^2\bigr)
\projdot2.4}{\bigl(\gamma_{12}^2-m_1^2 m_2^2\bigr)\projdot1.4}
\,, \\[0.8mm]
\xib'_3 \hspace{0.7mm}&\equiv\hspace{0.7mm}
  -\frac{m_4^2 \bigl( s_{34}-\gamma_{34}-m_4^2\bigr)
\projdot1.3}{\bigl(\gamma_{34}^2-m_3^2 m_4^2\bigr)\projdot1.4} \,, \qquad
&\xib'_4 \hspace{0.7mm}&\equiv\hspace{0.7mm}
  \frac{\gamma_{34} s_{34}-(\gamma_{34}+m_3^2) m_4^2}
                {\gamma_{34}^2-m_3^2 m_4^2}\,,
\end{alignedat}
\label{eq:def_of_xi_bar}
\end{equation}
\begin{equation}
\tau \hspace{0.7mm}\equiv\hspace{0.7mm}
  \frac{\spash{1^\proj}.{4^\proj}\spash{2^\proj}.{3^\proj}}
                   {\spash{2^\proj}.{4^\proj}\spash{1^\proj}.{3^\proj}}
= \frac{\spbsh{1^\proj}.{4^\proj}\spbsh{2^\proj}.{3^\proj}}
                    {\spbsh{2^\proj}.{4^\proj}\spbsh{1^\proj}.{3^\proj}}\,,
\label{taudef}
\end{equation}
where $m_i$ are the masses of the external momenta, $m_i^2=k_i^2$.  In
addition, we will make use of the following quantities not needed in
ref.~\cite{ExternalMasses}:
\begin{align}
\Delta \hspace{0.7mm}&\equiv\hspace{0.7mm} \Big((\bar\xi_1' + \bar\xi_2' ) \bar\xi_4' - (\bar\xi_1'  + \tau^2 \bar\xi_2')
\bar\xi_3' \Big)^2-4 \bar\xi_1' \bar\xi_2' (\tau \bar\xi_3' - \bar\xi_4')^2\,, \\[0.8mm]
z_\pm \hspace{0.7mm}&\equiv\hspace{0.7mm} \frac{1}{2 \xib'_1 (\tau \xib'_3 - \xib'_4)} \Big( (\xib'_1 + \xib'_2) \xib'_4
- (\xib'_1 + \tau^2 \xib'_2) \xib'_3 \hspace{1mm}\pm\hspace{1mm} \sqrt{\Delta} \Big) \label{eq:zplusminus}\,, \\[0.8mm]
\gamma_* \hspace{0.7mm}&\equiv\hspace{0.7mm} \frac{\gamma_{12}\gamma_{34}}
{32 k_1^\flat \cdot k_4^\flat \, (\gamma_{12}^2-m_1^2 m_2^2)(\gamma_{34}^2-m_3^2 m_4^2)} \,.
\end{align}

\section{Maximal Cuts of Double-Box Integrals}
\label{sec:four-mass_projectors}

Our aim is to determine the coefficients of the double-box master
integrals that appear in the basis expansion~(\ref{BasicEquation}) of a
two-loop quantity that may either be an amplitude, form factor, or
correlator. Without loss of generality, we refer to the two-loop quantity
as an amplitude.  The double-box integral topology is illustrated in
fig.~\ref{fig:DoubleBox}, and defines the internal momenta $p_j$. The
integral is defined in dimensional regularization with $D=4-2\epsilon$ as,
\begin{equation}
P_{2,2}^{**} [\Phi] \equiv \int_{\mathbb{R}^D \times \mathbb{R}^D}
\frac{d^D \ell_1}{(2\pi)^D} \frac{d^D \ell_2}{(2\pi)^D}
\frac{\Phi(k_1, k_2,k_3;\ell_1,\ell_2)}{\prod_{j=1}^7 p_j^2} \,,
\label{eq:def_of_DB}
\end{equation}
where $\Phi$ denotes an arbitrary polynomial in the external and internal
momenta.  We refer to it as a numerator insertion.
At one loop, all numerator insertions can be expressed as linear
combinations of propagator denominators, external invariants, and
parity-odd functions which vanish upon integration; but this is
no longer true at two loops and beyond.  At higher loops, some
polynomials $\Phi$ are irreducible.  Integrals with certain
irreducible-numerator insertions can be related to others using
IBP identities, but in general several will remain as master integrals.

We seek formulas for the double-box coefficients to
leading order in the dimensional regulator $\epsilon$ in terms of
purely tree-level input.
We begin by cutting all double-box propagators on both sides of
\eqn{BasicEquation}.  This immediately eliminates all integrals with
fewer than seven propagators, or with a different topology, as cutting
an absent propagator yields zero.

Heuristically, we may imagine using the Cutkosky rules, and simply
replacing the cut propagators by on-shell delta functions.  On the left-hand
side of the equation, we would then obtain,
\begin{equation}
A^{(2)}\Big|_{\rm cut} = (2\pi i)^7
\int \frac{d^4 \ell_1}{(2\pi)^4} \frac{d^4 \ell_2}{(2\pi)^4}
\prod_{j=1}^7 \delta(p_j^2) \prod_{v=1}^6A^{\tree}_{(v)}\,,
\label{eq:factorization}
\end{equation}
where $A^{\tree}_{(v)}$ denote the tree processes at each of the six
vertices of the diagram in fig.~\ref{fig:DoubleBox}, and $p_j$ denote
the momenta flowing through each of the propagators.
The cuts have also eliminated any potential infrared divergences,
so we can take the four-dimensional limit for the integrand.
On the right-hand side of \eqn{BasicEquation}, we would obtain
a sum over expressions of the form,
\begin{equation}
{\rm coefficient\ \/}\times  \, P_{2,2}^{**} [\Phi] \Big|_{\rm cut}  =
{\rm coefficient\ \/}\times \, (2\pi i)^7
\int \frac{d^4 \ell_1}{(2\pi)^4} \frac{d^4 \ell_2}{(2\pi)^4} \hspace{0.3mm}
\prod_{j=1}^7 \delta(p_j^2)\, \Phi \,.
\label{eq:formal_general_heptacut}
\end{equation}
If we interpret the expressions
in \eqns{eq:factorization}{eq:formal_general_heptacut} literally,
however, we face a problem.
The integrations in these equations
receive contributions only from regions of integration space where the loop
momenta solve the joint on-shell constraints,
\begin{equation}
p_j^2 = 0 \,, \hspace{9mm} j=1,\ldots,7 \label{eq:on-shell_constraints}\,.
\end{equation}
For generic external momenta, the solutions to these equations
are complex.  So long as the integrations are over real momenta
($\mathbb{R}^4 \times \mathbb{R}^4$), we simply get zero.  Equating
the two expressions will yield $0 = 0$, which is true but useless for
extracting the coefficient in \eqn{eq:formal_general_heptacut}.

Instead of thinking of the loop integrals as integrals over real
momenta, we can choose to think of them as integrals in complex
momenta, $\ell_i\in \mathbb{C}^4$, taken along contours comprising the
real slice, $\mathop{\rm Im}\nolimits\ell_i^\mu = 0$.  Changing the
contour then gives us an alternative way of imposing a delta-function
constraint, one that is valid for complex as well as for real
solutions.

\def\smallc{\varepsilon}
The utility of reinterpreting delta functions as contour integrals
was previously observed in the context of twistor-string
amplitudes~\cite{Roiban:2004yf,Vergu:2006np}, and is also standard
in more formal twistor-space expressions~\cite{MasonSkinner}.
In one dimension, we seek to localize an integral,
\begin{equation}
\int dq \hspace{0.8mm} \delta(q-q_0) h(q)\,
\hspace{1mm}=\hspace{1mm} h(q_0)
\end{equation}
even if $q_0$ becomes complex.  Cauchy's residue theorem gives us precisely
such a localization if we replace $\delta(q-q_0)$ by
$\frac{1}{2\pi i} \frac{1}{q - q_0}$, and take the integral to be a contour
integral along a small circle centered at $q_0$ in the complex $q$-plane.
Analogously, a product of delta functions can be defined as
a multidimensional contour integral,
\begin{equation}
(2\pi i)^n \int dq_1 \cdots dq_n \hspace{0.5mm} h(q_i)
\prod_{j=1}^n
\delta (q_j - q_{0j}) \stackrel{\rm def}{=} \int_{T_\smallc (q_0)}
 dq_1 \cdots dq_n \frac{h(q_i)}{\prod_{j=1}^n (q_j - q_{0j})}
\label{eq:delta_functions_as_contour_int}
\end{equation}
where the contour $T_\smallc (q_0)$ is now a torus encircling
the simultaneous solution of denominator equations.
For the simple form of the denominator here, the contour
will be a product of $n$ small circles ($\smallc\ll1$),
$T_\smallc (q_0) = C_\smallc (q_{01}) \times \cdots \times
C_\smallc (q_{0n})$,
each centered
at $q_{0j}$. The simultaneous solution of the denominator equations is
called a \emph{global pole\/}.
The question of what it means for a torus to encircle a global pole
is much more subtle in higher dimensions than for a contour to encircle
a point in one complex dimension; but the subtleties will play no role
in the present article.

There is one important respect in which the multidimensional contour
integrals behave differently from integrals over delta functions, namely
the transformation formula for changing variables.  Given a holomorphic
function $f = (f_1, \ldots, f_n) : \mathbb{C}^n \to \mathbb{C}^n$ with an
isolated zero\footnote{A function $f = (f_1, \ldots, f_n) : \mathbb{C}^n
  \to \mathbb{C}^n$ is said to have an isolated zero at $a \in
  \mathbb{C}^n$ iff by choosing a small enough neighborhood $U$ of $a$ one
  can ensure that it is has only a single zero in the neighborhood, so that
  $f^{-1}(0) \cap U = \{ a\}$.}  at $a \in \mathbb{C}^n$, the residue at
$a$ is computed by performing the integral over a toroidal
contour, whose general definition is $T_\smallc (a)
= \{ z \in \mathbb{C}^n : |f_i (z)| = \smallc_i,
\hspace{1mm} i = 1, \ldots, n \}$. This contour integral satisfies the
transformation formula
\begin{equation}
\frac{1}{(2\pi i)^n} \int_{T_\smallc (a)}
\frac{h(z) \hspace{0.4mm} dz_1 \wedge \cdots\wedge dz_n}
     {f_1 (z) \cdots f_n (z)} \hspace{1mm}=\hspace{1mm} \frac{h(a)}
{\det_{i,j} \frac{\partial f_i}{\partial z_j}} \,.
\label{eq:contour_integration_transf_formula}
\end{equation}
Unlike the conventional formula for a multidimensional real integral
over delta functions, it does not involve taking the absolute value of
the inverse Jacobian. This ensures that this factor is analytic in any
remaining variables on which it depends, so that further contour integrations
can be carried out.

We use multidimensional contour integrals to define generalized
discontinuity operators.  The GDOs for the
double box will be eightfold integrals taken over contours that are linear
combinations of basis contours.  Each basis contour encircles a single
global pole, and we will refer to global poles and their encircling
contours interchangeably.  Applying a GDO means changing the contour of the
integration from one over the real slice of
$\mathbb{C}^4\times\mathbb{C}^4$ to one over the GDO's associated contour.
We want seven of the eight contour integrations to correspond to the seven
on-shell constraints $p_j^2=0$; to do so, the contours must ultimately
encircle solutions to these constraints.  The integrands in
\eqns{eq:factorization}{eq:formal_general_heptacut} are left unchanged.
Imposing the seven constraints leaves one complex degree of freedom.  The
heptacut constraints thus define a Riemann surface in
$\mathbb{C}^4\times\mathbb{C}^4$.  As we will see below, this Riemann
surface contains a number of poles.  Their presence will allow us to freeze
the remaining degree of freedom, by choosing an appropriate contour of
integration for the corresponding localization variable.  Before discussing
the poles, however, we first review the structure of the Riemann surface.

\subsection{Kinematical Solutions, Jacobians and Global Poles}
\begin{figure}[th]
\begin{center}
\includegraphics[width=0.5\textwidth]{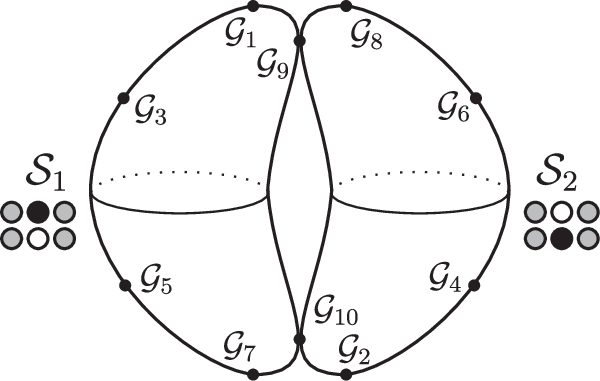}
\end{center}
\caption{\small A representation of the pinched torus solution space
  for the class~(a) heptacut kinematics, showing the two independent
  solutions $\Sol_i$, and the locations of the eight global poles
  $\Global_i$.  The small white, black and gray blobs indicate the
  pattern of chiral, antichiral and nonchiral kinematics,
  respectively, at the vertices of a double-box
  integral. Complex-conjugate pairs of poles are identified by
  reflection through the center of the torus.}
\label{ClassASolutionsFigure}
\end{figure}

As discussed in ref.~\cite{Caron-HuotLarsen},
the maximal-cut Riemann surface for
the double-box integral is a pinched torus, with the
number of pinches equal to twice the number of vertical double-box
rungs that attach to an on-shell massless three-point vertex.
An on-shell massless three-point vertex is
either chiral or anti-chiral, enforcing a two-fold branching of the
kinematical parametrization, and implying a pinching of the parameter
space. A more careful analysis of the kinematical solutions shows that chiral
vertices are (anti-)correlated across the vertical rungs of the
double-box integral. Hence, we classify the different types of pinches
by their effect on the vertical rungs.

In previous work~\cite{ExternalMasses}, we assigned double boxes to
one of three classes (a), (b), or (c), according to whether an
on-shell massless three-point vertex is connected to: (a) the middle
rung, (b) the middle rung and one outer rung, (c) all three vertical
rungs\footnote{$P_{2,2}$ or `flying-squirrel' integrals (in the
  notation of ref.~\cite{TwoLoopBasis}), with external legs attached
  to the middle vertices of the double box, would yield novel
  classes~\cite{Caron-HuotLarsen}, and are not treated here.}.  We
treated the two latter classes in ref.~\cite{ExternalMasses}.  Here,
we consider class~(a), corresponding to the four-mass double box,
illustrated in fig.~\ref{fig:DoubleBox}. In this class, the solutions
to the heptacut equations~(\ref{eq:on-shell_constraints}) form a
doubly-pinched torus, shown in fig.~\ref{ClassASolutionsFigure}.

Each lobe of the doubly-pinched torus corresponds to one of
two kinematical solutions, ${\cal S}_1$ and ${\cal S}_2$.
In terms of the loop momentum parametrization of
\eqn{SpinorParametrization}, both solutions have $\xi'_1=\xib'_1$,
$\xi'_4=\xib'_4$, $\xi_1 = \xi_4 = 1$, and the remaining four
variables $(\xi_2,\xi'_2,\xi_3,\xi'_3)$ take on the following values,
\begin{equation}
\begin{aligned}
\Sol_1:& \hspace{4mm}
\left(\frac{\xib'_2}{z}, \hspace{0.9mm} z, \hspace{0.9mm} -\frac{(z + \xib'_1/\tau) \tau \xib'_3}
{(z + \xib'_1) \xib'_4}, \hspace{0.9mm} -\frac{(z + \xib'_1) \xib'_4}
{(z + \xib'_1/ \tau)\tau} \right)\,,
\\
\Sol_2:& \hspace{4mm}
\left(z, \hspace{0.9mm} \frac{\xib'_2}z, \hspace{0.9mm} -\frac{z+1}{\tau z +1}, \hspace{0.9mm}
-\frac{(\tau z +1)\xib'_3}{z+1}\right)\,,
\end{aligned}
\end{equation}
where the $\xib'_i$ are defined in \eqn{eq:def_of_xi_bar}.  The
Jacobian that arises from changing the integration variables of
eq.~(\ref{eq:def_of_DB}) to the $\xi_i, \xi'_i, \zeta_i$ in
eqs.~(\ref{SpinorParametrization})--(\ref{BasicSpinorDefinitions}) and
subsequently
performing seven contour integrals, $\int d^4\ell_1 d^4 \ell_2
\prod_{j=1}^7 1/p_j^2
\longrightarrow \int dz\, \ji$, takes the generic form
\begin{equation}
\ji (z)\big|_{\mathcal{S}_i} \equiv
\left( \det_{\mu,i} \frac{\partial \ell_1^\mu}{\partial v_{1,i}} \right)
\left( \det_{\nu,j} \frac{\partial \ell_2^\nu}{\partial v_{2,j}} \right)
\left( \det_{i,j} \frac{\partial p_i^2}{\partial w_j} \right)^{-1}
= \frac{C_i}{(z - z_{i,1}) (z - z_{i,2})} \,,
\label{eq:Jacobian_formal}
\end{equation}
where in the first equality
$v_{j,1} = \zeta_j$, $v_{1,2} = \xi'_1$, $v_{2,2} = \xi'_4$,
$v_{1,3} = \xi_2$, $v_{2,3} = \xi_3$, $v_{1,4} = \xi'_2$,
and $v_{2,4} = \xi'_3$, and $w_j$ are the seven variables frozen
by the contour integrations.
In the second equality, $(z_{i,1}, z_{i,2})$ are the local coordinates of the
intersection with the neighboring solution(s),
\begin{equation}
\begin{aligned}
\mathcal{S}_i \big|_{z=z_{i,1}} \hspace{2mm}
  &\in \hspace{2mm} \mathcal{S}_{i-1} \cap \mathcal{S}_i \\
\mathcal{S}_i \big|_{z=z_{i,2}} \hspace{2mm}
  &\in \hspace{2mm} \mathcal{S}_i \cap \mathcal{S}_{i+1} \,.
\end{aligned}
\end{equation}
More generally, the Jacobian evaluated on a Riemann sphere will
always be a product of simple-pole factors associated with the
pinching points (also known as nodal points) on the sphere.

As mentioned above, the heptacut of the double-box integral arises
from performing seven of the eight contour integrals, and yields a
Riemann surface given by the solution to the joint on-shell
constraints (\ref{eq:on-shell_constraints}).  We are left with a
single complex degree of freedom (or localization variable) $z$, and
the freedom to choose a contour for its integration.  In order to
localize the integrand completely, we should have this last contour
encircle a pole in $z$.  As in classes~(b) and~(c) treated in
ref.~\cite{ExternalMasses}, such poles can arise from two sources: the
Jacobian factor~(\ref{eq:Jacobian_formal}), or from the numerator
insertions $\Phi$ in \eqn{eq:def_of_DB}, which introduce an additional
dependence on $z$.

\def\GRes#1{\mathop{\mathrm{Res}}_{\,\,\,\mathcal{G}_{#1}}}
\def\GResT#1{\mathop{\mathrm{Res}}_{\,\mathcal{G}_{#1}}}

The Jacobian poles are the pinching points $\mathcal{G}_{9,10}$ in
fig.~\ref{ClassASolutionsFigure}.  Because these points are shared between
different on-shell solutions, one must decide on a convention for the
sphere on which the corresponding residue is to be evaluated. We adopt the
convention of computing the residue on the sphere located on the
anti-clockwise side of the Riemann surface. In
fig.~\ref{ClassASolutionsFigure}, for example, the residue at
$\mathcal{G}_9$ should be evaluated on $\mathcal{S}_1$; and the residue at
$\mathcal{G}_{10}$ on $\mathcal{S}_2$.  Furthermore, we choose the
orientations on each Riemann sphere such that for any global pole
$\Global_k \in \Sol_i \cap \Sol_j$, the residues evaluated on spheres
$\Sol_i$ and $\Sol_j$ are equal in magnitude but opposite in sign. That is,
for an arbitrary function~$f$ of the loop momenta one has,
\begin{equation}
\mathop{\mathrm{Res}}_{\mathcal{S}_i \cap \mathcal{S}_{i+1}} \ji (z)
f(\ell_1 (z), \ell_2 (z)) \Big|_{\mathcal{S}_i}
\hspace{1.5mm}=\hspace{1.5mm} -
\mathop{\mathrm{Res}}_{\mathcal{S}_i \cap \mathcal{S}_{i+1}} \ji (z)
f(\ell_1 (z), \ell_2 (z)) \Big|_{\mathcal{S}_{i+1}} \, ,
\label{eq:orientation_conventions}
\end{equation}
in agreement with the conventions of
refs.~\cite{Caron-HuotLarsen,ExternalMasses}.  Other
choices of conventions are possible, but all will lead to the same
final expressions for the two-loop integral coefficients.

The class~(a) Jacobian, as defined in eq.~(\ref{eq:Jacobian_formal}),
takes the form
\begin{equation}
\ji =
\left\{\begin{array}{ll}
\displaystyle
\ji^{(a,1)} \equiv \frac{\xib'_1}{\big(z - \xib'_1 z_+ \big)
                                          \big(z - \xib'_1 z_- \big)} \,,
&{\rm for\ solution\/}~ \Sol_1\,,\\[4mm]
\displaystyle
\ji^{(a,2)} \equiv
\frac{1}{(z - z_+)(z - z_-)}\,,
&{\rm for\ solution\/}~ \Sol_2\,,
\end{array}\right. \label{eq:ClassA_Jacobian}
\end{equation}
after the convenient rescaling $\ji \hspace{0.8mm}\longrightarrow\hspace{0.8mm}
\frac{\xib'_1 (\xib'_4 - \tau \xib'_3)}{\gamma_*} \ji$, in analogy
with eqs.~(4.7,4.8) of ref.~\cite{ExternalMasses}.
(The rescaling of course leaves the final formul\ae{} for integral coefficients unchanged.)
Note that $z_+\,z_- = \xib'_2/\xib'_1$.

In addition to the Jacobian poles,
we may choose the contour for the remaining post-heptacut degree of freedom
to encircle any of
the points on the Riemann surface where a loop momentum becomes infinite.
At such points, numerator insertions $\Phi(p_i)$ have a pole in $z$.
We will refer to these points,
shown as punctures on the spheres in fig.~\ref{ClassASolutionsFigure},
 as insertion poles.
There are eight such poles, so that altogther we have ten global
poles which the $z$ contour integral may encircle,
located at the following values
of $(\xi_2, \xib'_2, \xi_3, \xib'_3)$:
\def\xsp{\hspace{0.5mm}}
\def\interc{\hspace{3mm}}
\def\chiral{\raise-0.5mm\hbox{\huge  $\bullet$}}
\def\antichiral{\raise-0.5mm\hbox{\huge $\circ$}}
\begin{equation}
\begin{alignedat}{2}
\Global_1\!: & \biggl( -\frac{\xib'_2}{\xib'_1},\xsp  -\xib'_1,
                       \xsp \infty,\xsp 0 \biggr)\,, &\interc
\Global_2\!: & \bigg( -1,\xsp -\xib'_2,\xsp 0,\xsp \infty \bigg)\,,
   \\[1mm]
\Global_3\!: & \bigg( 0,\xsp \infty,\xsp -\frac{\tau \xib'_3}{\xib'_4},
                      \xsp -\frac{\xib'_4}{\tau} \bigg)\,, &\interc
\Global_4\!: & \bigg( \infty,\xsp 0,\xsp -\frac{1}{\tau},
                    \xsp -\tau \xib'_3 \bigg)\,, \\[1mm]
\Global_5\!: & \bigg( -\frac{\tau \xib'_2}{\xib'_1},
                    \xsp -\frac{\xib'_1}{\tau},\xsp 0,\xsp \infty \bigg)\,,
            &\interc
\Global_6\!: & \bigg( -\frac{1}{\tau},\xsp -\tau \xib'_2,
                    \xsp \infty,\xsp 0 \bigg)\,,\\[1mm]
\Global_7\!: & \bigg( \infty,\xsp 0,\xsp -\frac{\xib'_3}{\xib'_4},
                    \xsp -\xib'_4 \bigg)\,, &\interc
\Global_8\!: & \bigg( 0,\xsp \infty, \xsp -1,
                    \xsp -\xib'_3 \bigg)\,, \\
\Global_9\!: & \bigg(z_+,\xsp \frac{\xib'_2}{z_+},
                     \xsp -\frac{(z_+ \xib'_1 + \tau \xib'_2) \xib'_3}
                                {(z_+ \xib'_1 + \xib'_2) \xib'_4},
                     \xsp -\frac{(z_+ \xib'_1 + \xib'_2) \xib'_4}
                                {z_+ \xib'_1 + \tau \xib'_2} \bigg),
                       &
\Global_{10}\!: & \bigg(z_-,\xsp \frac{\xib'_2}{z_-},
                        \xsp -\frac{1+z_-}{1+\tau z_-},
                        \xsp-\frac{(1+\tau z_-)\xib'_3}{1+z_-}\bigg)\,.
\hspace{-5mm}
\end{alignedat}\hspace{-5mm}
\label{ClassGlobalPoles}
\end{equation}
In the above labeling, the poles $(\mathcal{G}_{2j-1},
\mathcal{G}_{2j}), \hspace{1mm} j=1,\ldots,7$ form parity-conjugate
pairs.  Because parity amounts to swapping chiralities
\chiral{} $\longleftrightarrow$ \antichiral, thereby rotating
fig.~\ref{ClassASolutionsFigure} by an angle $\pi$, parity-conjugate
pairs always appear antipodally in the figure.  We note that at the
pinching points $\mathcal{G}_{9,10}$, the loop momentum flowing
through the middle rung of the double box becomes soft, $p_4
\rightarrow 0$. At the remaining global poles, either the left or
right loop momentum goes to infinity in a particular direction.  (See
the appendix for a more detailed discussion.)

Let us denote a contour consisting of a small circle around
$\mathcal{G}_j$ by $C_j$.  The set of circles around all of the ten
poles in fig.~\ref{ClassASolutionsFigure} forms an overcomplete basis
of contours for GDOs, and equivalently an overcomplete basis for
homology.  On each sphere we can use the fact that all residues sum to
zero to eliminate any one contour $C_j$ in favor of the remaining
ones.  This is not sufficient, as we must impose additional
consistency constraints on the linear combination of contours by which
every GDO acts.  We discuss these below.  Retaining all contours
instead of choosing a linearly-independent subset does have the
advantage of making manifest certain discrete symmetries, clarifying
the structure of the additional consistency constraints.  We examine
this issue in more detail in sec.~\ref{sec:Feynman_varieties}.

The truncation to a linearly independent homology basis can be achieved
simply by setting the coefficients of certain contours to zero in every
GDO.  Not all truncations will lead to a valid basis, however; a basis must
necessarily contain a contour encircling at least one of the pinching
points $\mathcal{G}_{9,10}$. To understand why, consider the sum of all
residues on $\mathcal{S}_1$ plus the sum of all residues on
$\mathcal{S}_2$. Both sums are zero by Cauchy's theorem, and this sum is
therefore zero. On the other hand, the sum equals that over the insertion
poles alone, $\sum_{i=1}^8 \GRes{i}$, because
the contributions from the pinching points cancel owing to
eq.~(\ref{eq:orientation_conventions}). We thus conclude that the residues
at the insertion poles always sum to zero, and the set of contours
encircling insertion poles alone does not constitute a complete homology
basis on $\mathcal{S}_1 \cup \mathcal{S}_2$.

\subsection{Master Contours -- General Four-Mass Kinematics}
\label{sec:master_contours}

Generalized discontinuity operators for the planar double box are given
as eightfold contour integrals, which factor into a sevenfold contour integral localizing
the integrand onto the heptacut solution surface --- the joint solution of the on-shell equations for all
seven propagator momenta.  The last contour integral is now a contour integral on
that Riemann surface.
The contour cannot be chosen arbitrarily, however.  It
is subject to the consistency requirement that it yield a vanishing
integration for any function that integrates to zero
on the original contour of integration for the Feynman
integral, $\mathbb{R}^D \times \mathbb{R}^D$. This ensures that two integrals
which are equal, for example by virtue of non-trivial integral
relations, have the same generalized cut,
\begin{equation}
\mathrm{Int}_1 \hspace{1mm}=\hspace{1mm} \mathrm{Int}_2 \hspace{7mm}
\Longrightarrow \hspace{7mm} \GenDisc(\mathrm{Int}_1)
\hspace{1mm}=\hspace{1mm} \GenDisc(\mathrm{Int}_2)\,.
\end{equation}
Examples of terms which integrate to zero
on
$\mathbb{R}^D \times \mathbb{R}^D$ include parity-odd
terms and total derivatives used in the integration-by-parts identities
to reexpress a large set of formally-irreducible integrals in terms of
linearly independent master integrals.

We may write a general contour for a GDO as follows,
\begin{equation}
\sum_i \omega_i C_i
\label{GeneralContour}
\end{equation}
where the $\omega_i$ are complex coefficients, and where the sum is
taken over a linearly independent homology basis (or over an
overcomplete one).  For double-box integrals belonging to class~(a), it
turns out that consistency with IBP relations imposes no constraints
on the contour, and hence no constraints on the $\omega_i$.  On the
other hand, the vanishing integration of (parity-odd) Levi-Civita numerator
insertions --- such as $\varepsilon(\ell_1,k_1,k_2,k_4)$
--- results in the following constraints on the coefficients
$\omega_i$:
\begin{equation}
\begin{array}{rl}
2\omega_1 - 2\omega_2 - \omega_9 + \omega_{10}  \hspace{1mm}=&\hspace{-0.5mm}  0 \,\phantom{.} \\
2\omega_3 - 2\omega_4 - \omega_9 + \omega_{10}  \hspace{1mm}=&\hspace{-0.5mm}  0 \,\phantom{.} \\
2\omega_5 - 2\omega_6 - \omega_9 + \omega_{10}  \hspace{1mm}=&\hspace{-0.5mm}  0 \,\phantom{.} \\
2\omega_7 - 2\omega_8 - \omega_9 + \omega_{10}  \hspace{1mm}=&\hspace{-0.5mm}  0 \,.
\end{array}\label{eq:ClassA_LeviCivita_constraints}
\end{equation}
In class (a), there are four linearly independent double-box
integrals which we may choose to be,
\begin{equation}
(I_1, I_2, I_3, I_4) \hspace{0.8mm}=\hspace{0.8mm} \big( P_{2,2}^{**} [1], \hspace{1mm} P_{2,2}^{**} [\ell_1 \cdot k_4], \hspace{1mm}
P_{2,2}^{**} [\ell_2 \cdot k_1], \hspace{1mm} P_{2,2}^{**} [(\ell_1 \cdot k_4)(\ell_2 \cdot k_1)] \big) \,.
\label{eq:master_integrals}
\end{equation}
The residues at the global poles $(\mathcal{G}_1, \ldots, \mathcal{G}_{10})$
of these integrals are as follows,
\begin{equation}
\begin{array}{rl}
\GRes{i} P_{2,2}^{**} [1]                                     \hspace{2mm}=&\hspace{-1mm} \frac{\bar\xi_1' (\bar\xi_4' - \tau \bar\xi_3')}{\sqrt{\Delta}}
                                                                         \big(0, 0, 0, 0, 0, 0, 0, 0, 1, 1\big) \\[3mm]
\GRes{i} P_{2,2}^{**} [\ell_1 \cdot k_4]                      \hspace{2mm}=&\hspace{-1mm} \frac{\bar\xi_1' (\bar\xi_4' - \tau \bar\xi_3')}{\sqrt{\Delta}}
                                                                         \big(0, 0, r_4, r_4, 0, 0, -r_4, -r_4, r_2, r_2\big) \\[3mm]
\GRes{i} P_{2,2}^{**} [\ell_2 \cdot k_1]                      \hspace{2mm}=&\hspace{-1mm} \frac{\bar\xi_1' (\bar\xi_4' - \tau \bar\xi_3')}{\sqrt{\Delta}}
                                                                         \big(r_3, r_3, 0, 0, -r_3, -r_3, 0, 0, r_1, r_1\big) \\[3mm]
\GRes{i} P_{2,2}^{**} [(\ell_1 \cdot k_4)(\ell_2 \cdot k_1)]  \hspace{2mm}=&\hspace{-1mm} \frac{\bar\xi_1' (\bar\xi_4' - \tau \bar\xi_3')}{\sqrt{\Delta}}
                                                                         \big(r_6, \hspace{0.6mm} r_6, \hspace{0.6mm} r_7, \hspace{0.6mm} r_7, \hspace{0.6mm} r_8, \hspace{0.6mm} r_8, \hspace{0.6mm} r_5, \hspace{0.6mm} r_5, \hspace{0.6mm} r_1 r_2, \hspace{0.6mm} r_1 r_2\big) \,,
\end{array}\label{eq:master_integrals_residues}
\end{equation}
where the $r_i$ are given by\footnote{Note that as expected
$\sum_{i=1}^8 \GRes{i} I_j =0$ for all
four master integrals as
\begin{equation}
r_5 + r_6 + r_7 + r_8 = 0 \,,
\end{equation}
consistent with the discussion below eq.~(\ref{ClassGlobalPoles}).},
\begin{equation}
\begin{array}{rll}
r_1 &=& \displaystyle
-\frac{1}{2} \hspace{-0.7mm} \left(2 \bar\xi_2' k^\flat_1 \cdot k^\flat_2 \frac{k^\flat_1 \cdot k^\flat_4}{k^\flat_2 \cdot k^\flat_4} + \bar\xi_1' m_1^2 \right) \,, \\[2.8mm]
r_2 &=& \displaystyle
-\frac{1}{2} \hspace{-0.7mm} \left(2\bar\xi_3' k^\flat_3 \cdot k^\flat_4 \frac{k^\flat_1 \cdot k^\flat_4}{k^\flat_1 \cdot k^\flat_3}  + \bar\xi_4' m_4^2 \right) \,, \\[2.8mm]
r_3 &=& \displaystyle
 \frac{1}{2} \frac{\spb{1^\flat}.{4^\flat} }{\spb{1^\flat}.{3^\flat}}  \frac{\sqrt{\Delta}}{\tau \bar\xi_2' - \bar\xi_1'} \langle 4^\flat|\slashed{k}_1 | 3^\flat] \,, \\[2.8mm]
r_4 &=& \displaystyle
 \frac{1}{2} \frac{\spb{1^\flat}.{4^\flat}}{\spb{2^\flat}.{4^\flat}}  \frac{\sqrt{\Delta}}{\tau \bar\xi_3' - \bar\xi_4'} \langle 1^\flat|\slashed{k}_4 | 2^\flat] \,, \\[2.8mm]
r_5 &=& \displaystyle
 \frac{1}{4} m_1^2 \sqrt{\Delta} \frac{\spa{3^\flat}.{4^\flat} \spb{1^\flat}.{4^\flat}}{\spa{1^\flat}.{3^\flat} \spb{1^\flat}.{2^\flat}} \langle 1^\flat|\slashed{k}_4 | 2^\flat] \,, \\[2.8mm]
r_6 &=& \displaystyle
 \frac{1}{4} m_4^2 \sqrt{\Delta} \frac{\spa{1^\flat}.{2^\flat} \spb{1^\flat}.{4^\flat}}{\spa{4^\flat}.{2^\flat} \spb{3^\flat}.{4^\flat}} \langle 4^\flat|\slashed{k}_1 | 3^\flat] \,, \\[2.8mm]
r_7 &=& \displaystyle
 \frac{1}{2} k^\flat_1 \cdot k^\flat_4 \sqrt{\Delta} \frac{\spb{3^\flat}.{4^\flat} \spb{1^\flat}.{2^\flat}}{\spb{2^\flat}.{3^\flat} \spb{2^\flat}.{4^\flat}} \langle 1^\flat|\slashed{k}_4 | 2^\flat] \,, \\[2.8mm]
r_8 &=& \displaystyle
 \frac{1}{2} k^\flat_1 \cdot k^\flat_4 \sqrt{\Delta} \frac{ \spb{1^\flat}.{2^\flat} \spb{3^\flat}.{4^\flat}}{\spb{3^\flat}.{2^\flat} \spb{1^\flat}.{3^\flat}} \langle 4^\flat|\slashed{k}_1 | 3^\flat] \,.
\end{array}\label{eq:definition_of_r_i}
\end{equation}
At this point, let us choose a linearly independent homology basis for
$\mathcal{S}_1 \cup \mathcal{S}_2$ consisting of the small circles
$C_{3,\ldots,10}$ encircling the global poles
$(\mathcal{G}_i)_{i=3,\ldots,10}$.  This leaves us with eight coefficients,
one for each $C_j$, subject to the four constraints in
eq.~(\ref{eq:ClassA_LeviCivita_constraints}).  Overall, we are left with
four independent coefficients, the same as the number of class~(a)
double-box master integrals, as given in  \eqn{eq:master_integrals}.

We can solve for these independent coefficients, finding a unique solution
which yields one when applied to one of the master integrals, and zero to
the others.  There are four such solutions, one for each master integral.
We refer to the contours as \emph{projectors} or \emph{master contours},
and to the operations of replacing the original integration contour by one
of these contours and performing the contour integrals as the GDO.  Each
GDO uniquely extracts the coefficient of one of the master integrals in the
basis decomposition (\ref{BasicEquation}) of the two-loop amplitude.  Using
the homology basis specified above, the master contours,
\begin{equation}
\Gamma_j =
\boldsymbol{\Omega}_j\cdot \mathbf{C} = \sum_{i=3}^{10} \omega_{j,i} C_i\,,
\end{equation}
associated with the master integrals $I_j$ in
eq.~(\ref{eq:master_integrals}) are given by the following coefficients,
\begin{equation}
\begin{aligned}
I_1 &: \hspace{3mm} \boldsymbol{\Omega}_1 \hspace{1mm}=\hspace{1mm} \frac{\sqrt{\Delta}}{\bar\xi_1' (\bar\xi_4' - \tau \bar\xi_3')}
\left( -\frac{r_2 r_3 r_5 + r_1 r_4 (r_2 r_3 + r_8)}
{2r_3 r_4 (r_5 + r_7)},\hspace{0.7mm} -\frac{r_2 r_3 r_5 + r_1 r_4 (r_2 r_3 + r_8)}
{2r_3 r_4 (r_5 + r_7)},\hspace{0.7mm} \frac{r_1}{2r_3},\hspace{0.7mm} \frac{r_1}{2r_3}, \right.  \\
&\hspace{43.6mm}\left. \frac{r_2 r_3 r_7 - r_1 r_4 (r_2 r_3 + r_8)} {2r_3 r_4 (r_5 + r_7)},\hspace{0.7mm}
\frac{r_2 r_3 r_7 - r_1 r_4 (r_2 r_3 + r_8)} {2r_3 r_4 (r_5 + r_7)},\hspace{0.7mm}\frac{1}{2},\hspace{0.7mm}
\frac{1}{2} \right)  \\[2mm]
I_2 &: \hspace{3mm} \boldsymbol{\Omega}_2 \hspace{1mm}=\hspace{1mm} \frac{\sqrt{\Delta}}{\bar\xi_1' (\bar\xi_4' - \tau \bar\xi_3')} \frac{1}{2r_4 (r_5 + r_7)}
\left( r_5,\hspace{0.6mm} r_5,\hspace{0.6mm} 0,\hspace{0.6mm} 0,\hspace{0.6mm}
-r_7,\hspace{0.6mm} -r_7,\hspace{0.6mm} 0,\hspace{0.6mm} 0 \right)  \\[2mm]
I_3 &: \hspace{3mm} \boldsymbol{\Omega}_3 \hspace{1mm}=\hspace{1mm} \frac{\sqrt{\Delta}}{\bar\xi_1' (\bar\xi_4' - \tau \bar\xi_3')}
\frac{1}{2r_3 (r_5 + r_7)}
\left( r_8, \hspace{0.2mm} r_8,\hspace{0.2mm} -r_5 -r_7,\hspace{0.2mm} -r_5 -r_7,\hspace{0.2mm}
r_8,\hspace{0.2mm} r_8,\hspace{0.2mm}
0,\hspace{0.2mm} 0 \right)  \\[2mm]
I_4 &: \hspace{3mm} \boldsymbol{\Omega}_4 \hspace{1mm}=\hspace{1mm} \frac{\sqrt{\Delta}}{\bar\xi_1' (\bar\xi_4' - \tau \bar\xi_3')} \frac{1}{2(r_5 + r_7)}
\left( 1,\hspace{0.6mm} 1,\hspace{0.6mm} 0,\hspace{0.6mm} 0,\hspace{0.6mm} 1,
\hspace{0.6mm} 1,\hspace{0.6mm} 0,\hspace{0.6mm} 0 \right) \,.
\end{aligned}
\hspace{-10mm}
\label{eq:master_contour}
\end{equation}
(In refs.~\cite{MaximalTwoLoopUnitarity,ExternalMasses}, these
coefficients were labeled $P_j$.)
In terms of these contours, the double-box coefficients in the basis
expansion (\ref{BasicEquation}) are
given by the following formula,
\begin{equation}
{\rm coefficient}_j \hspace{0.8mm}=\hspace{0.8mm} \oint_{\Gamma_j} dz
\hspace{0.6mm} \ji^{(a,i)} \prod_{v=1}^6 A_{(v)}^\mathrm{tree} (z)
\end{equation}
where the Jacobian of \eqn{eq:ClassA_Jacobian}, $\ji^{(a,i)}$, is evaluated on
solution $\mathcal{S}_1$ or $\mathcal{S}_2$, according to the location of
the poles encircled by $\Gamma_j$.

\subsection{Master Contours -- Equal-Mass Case}

In the special-kinematics situation where
$k_1^2 = k_4^2 = m_1^2$ and $k_2^2 = k_3^2 = m_2^2$, all Lorentz scalars
are invariant under $(k_1, k_2) \longleftrightarrow (k_4, k_3)$, and one
additional integral identity arises,
\begin{equation}
P_{2,2}^{**} [\ell_1 \cdot k_4] \hspace{0.8mm}=\hspace{0.8mm}
P_{2,2}^{**} [\ell_2 \cdot k_1] \,. \label{eq:equal-mass_identity}
\end{equation}
We note that this identity does not arise as an IBP relation.
This identity may be used to eliminate one of the integrals in
eq.~(\ref{eq:master_integrals}), leaving us with three independent
master integrals whose associated master contours we provide below.
(Other special cases, such as $k_1^2=k_3^2 = m_1^2$ and $k_2^2=k_4^2=m_2^2$,
can be treated similarly.)

With equal-mass kinematics, all ten global poles in
\eqn{ClassGlobalPoles} remain distinct. In terms of the quantities,
\begin{equation}
\begin{aligned}
\rho_1 &= -\frac{m_1^2}{2} \,, \\[2.3mm]
\rho_2 &= -\frac{\gamma_{12} + m_2^2}{s_{12}} \frac{k_1^\flat \cdot k_4^\flat \sqrt{\Delta}}
{\bar\xi_1'} \,, \\[2.3mm]
\rho_3 &= \frac{m_1^2}{2} \frac{m_2^2 + \gamma_{12}}{m_1^2 + \gamma_{12}} \frac{k_1^\flat \cdot k_4^\flat \sqrt{\Delta}}
{\bar\xi_1'} \,, \\[2.3mm]
\rho_4 &= \frac{\gamma_{12}}{2} \frac{k_1^\flat \cdot k_4^\flat \sqrt{\Delta}}
{\bar\xi_1'} \,,
\end{aligned}
\end{equation}
the residues at the global poles $(\mathcal{G}_1, \ldots, \mathcal{G}_{10})$
of the four integrals in eq.~(\ref{eq:master_integrals}) take the form,
\begin{equation}
\begin{aligned}
\GRes{i} P_{2,2}^{**} [1]                   \hspace{2mm}=& \frac{\bar\xi_1'}{\sqrt{\Delta}} \frac{s_{12}}{\gamma_{12} + m_1^2}
                                                                         \big(0, 0, 0, 0, 0, 0, 0, 0, 1, 1\big)\,, \\[1mm]
\GRes{i} P_{2,2}^{**} [\ell_1 \cdot k_4]    \hspace{2mm}=& \frac{\bar\xi_1'}{\sqrt{\Delta}} \frac{s_{12}}{\gamma_{12} + m_1^2}
                                                                         \big(0, 0, \rho_2, \rho_2, 0, 0, -\rho_2, -\rho_2, \rho_1, \rho_1\big)\,, \\[1mm]
\GRes{i} P_{2,2}^{**} [\ell_2 \cdot k_1]    \hspace{2mm}=& \frac{\bar\xi_1'}{\sqrt{\Delta}} \frac{s_{12}}{\gamma_{12} + m_1^2}
                                                                         \big(\rho_2, \rho_2, 0, 0, -\rho_2, -\rho_2, 0, 0, \rho_1, \rho_1\big)\,, \\[1mm]
\GRes{i} P_{2,2}^{**} [(\ell_1 \cdot k_4)(\ell_2 \cdot k_1)]
\hspace{2mm}=& \frac{\bar\xi_1'}{\sqrt{\Delta}} \frac{s_{12}}{\gamma_{12} + m_1^2}
                                                                         \big(\rho_3, \hspace{0.6mm} \rho_3, \hspace{0.6mm} \rho_4, \hspace{0.6mm} \rho_4, \hspace{0.6mm} -\rho_4, \hspace{0.6mm} -\rho_4, \hspace{0.6mm} -\rho_3, \hspace{0.6mm} -\rho_3, \hspace{0.6mm}
                                                                         \rho_1^2, \hspace{0.6mm} \rho_1^2 \big) \,.
\end{aligned}
\end{equation}
We may use the identity (\ref{eq:equal-mass_identity}) to eliminate one of the
integrals in eq.~(\ref{eq:master_integrals}), leaving us with
the master integrals
\begin{equation}
(I_1, I_2, I_3) \hspace{0.8mm}=\hspace{0.8mm} \big( P_{2,2}^{**} [1], \hspace{1mm} P_{2,2}^{**} [\ell_1 \cdot k_4], \hspace{1mm}
P_{2,2}^{**} [(\ell_1 \cdot k_4)(\ell_2 \cdot k_1)] \big) \,.
\label{eq:equal-mass_master_integrals}
\end{equation}
The requirement that the heptacut contour respect the identity~(\ref{eq:equal-mass_identity})
yields the contour constraint
\begin{equation}
\omega_1 + \omega_2 - \omega_3 - \omega_4 - \omega_5 - \omega_6
+ \omega_7 + \omega_8 = 0 \,. \label{eq:equal-mass_contour_constraint}
\end{equation}
In terms of the basis of homology specified in sec.~\ref{sec:master_contours},
the master contours $\Gamma_j$ associated
with the master integrals in eq.~(\ref{eq:equal-mass_master_integrals}) take
the form,
\begin{equation}
\begin{aligned}
I_1 &: \hspace{3mm} \boldsymbol{\Omega}_1 \hspace{1mm}=\hspace{1mm} \frac{\sqrt{\Delta}}{2\bar\xi_1'}
\frac{\gamma_{12} + m_1^2}{s_{12}} \frac{\rho_1}{\rho_2}
\left( \frac{\rho_1 \rho_2 - \rho_3 - \rho_4}{\rho_3 - \rho_4},\hspace{0.7mm}
\frac{\rho_1 \rho_2 - \rho_3 - \rho_4}{\rho_3 - \rho_4},\hspace{0.7mm} 1,
\hspace{0.7mm} 1, \right.  \\
&\hspace{60mm}\left. \frac{\rho_1 \rho_2 - 2\rho_4}{\rho_3 - \rho_4},
\hspace{0.7mm} \frac{\rho_1 \rho_2 - 2\rho_4}{\rho_3 - \rho_4},
\hspace{0.7mm} \frac{\rho_2}{\rho_1}, \hspace{0.7mm} \frac{\rho_2}{\rho_1} \right)\,, \\[1mm]
I_2 &: \hspace{3mm} \boldsymbol{\Omega}_2 \hspace{1mm}=\hspace{1mm}
\frac{\sqrt{\Delta}}{2\bar\xi_1'}
\frac{\gamma_{12} + m_1^2}{s_{12}} \frac{1}{\rho_2}
\left( \frac{\rho_3 + \rho_4}{\rho_3 - \rho_4}, \hspace{0.7mm}
\frac{\rho_3 + \rho_4}{\rho_3 - \rho_4}, \hspace{0.7mm} -1,
\hspace{0.7mm} -1, \frac{2\rho_4}{\rho_3 - \rho_4}, \hspace{0.7mm}
\frac{2\rho_4}{\rho_3 - \rho_4}, \hspace{0.7mm} 0, \hspace{0.7mm} 0
\right)\,, \\[1mm]
I_3 &: \hspace{3mm} \boldsymbol{\Omega}_3 \hspace{1mm}=\hspace{1mm}
-\frac{\sqrt{\Delta}}{2\bar\xi_1'}
\frac{\gamma_{12} + m_1^2}{s_{12}} \frac{1}{\rho_3 - \rho_4}
\left( 1,\hspace{0.6mm} 1,\hspace{0.6mm} 0,\hspace{0.6mm} 0,\hspace{0.6mm} 1,
\hspace{0.6mm} 1,\hspace{0.6mm} 0,\hspace{0.6mm} 0 \right) \,.
\end{aligned}
\end{equation}
One is not obliged to make use of the integral identity
(\ref{eq:equal-mass_identity}) and enforce the ensuing contour constraint
(\ref{eq:equal-mass_contour_constraint}); one could equally well expand the
equal-mass amplitude in terms of the slightly overcomplete basis in
eq.~(\ref{eq:master_integrals}), with the associated master contours given
in \eqn{eq:master_contour}.  Indeed,
since the energies of heavy particles follow a Breit-Wigner distribution,
an amplitude involving four massive vector bosons (e.g., $WZ \to WZ$)
will typically be required only for unequal masses; only when taking
the on-shell approximation would the equal-mass case arise.

\section{Varieties Arising from Feynman Graphs}
\label{sec:Feynman_varieties}

In this section we discuss the heptacut of the planar double-box integral, putting some
of the observations in sec.~\ref{sec:four-mass_projectors}
into the broader context of algebraic geometry.

\begin{figure}[th]
\begin{center}
\includegraphics[width=0.95\textwidth]{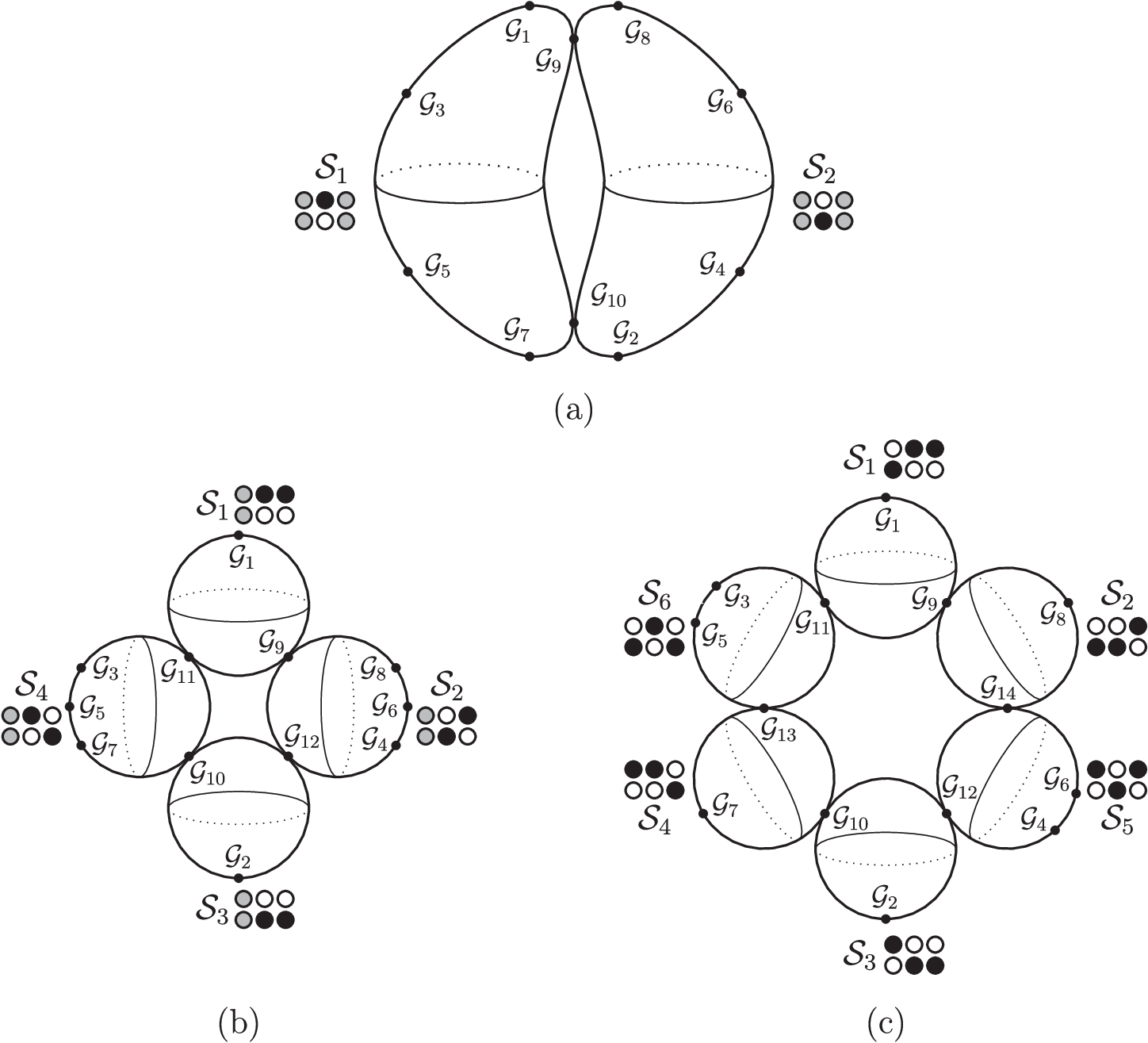}
\end{center}
{\vskip -3mm}
\caption{\small Representations of the solution space for the
class~(a), (b) and (c) heptacut equations, showing the independent solutions $\Sol_i$,
and the locations of the global poles $\Global_j$.}
\label{ClassABCSolutionsFigure}
\end{figure}

On-shell constraints are polynomial equations.  Accordingly,
their simultaneous solution defines an algebraic variety.
Ref.~\cite{Caron-HuotLarsen} observed that the variety corresponding
to setting all seven propagator momenta of the planar double box on-shell is a pinched torus, with
the number of pinches equal to twice the number of double-box rungs that end
on at least one three-point vertex.   As mentioned in the previous section
and in ref.~\cite{ExternalMasses}, we denote integrals
having one, two, or three such rungs as forming
classes~(a), (b), and (c). The respective pinched tori --- \emph{nodal elliptic curves}, in the
language of mathematicians --- are illustrated
in figs.~\ref{ClassABCSolutionsFigure}(a), \ref{ClassABCSolutionsFigure}(b),
and \ref{ClassABCSolutionsFigure}(c).

The components of the pinched tori are Riemann spheres. These spheres
are associated with distinct solutions to the joint on-shell
constraints (\ref{eq:on-shell_constraints}) and are characterized by
the distribution of chiralities (\chiral{} or \antichiral{})
at the vertices of the double-box graph. The fact that the number of
pinches is always even is a reflection of the fact that the on-shell
solutions always come in parity-conjugate pairs. At a pinching point,
there is exactly one double-box rung whose momentum becomes collinear
with the massless external momenta connected to the rung. For the
original uncut double-box integral, such regions of the loop momentum
integration typically produce infrared divergences, and the pinches
can therefore roughly be thought of as remnants of the original IR
divergences. In addition, the pinched tori contain a number of
insertion points (for example, in
fig.~\ref{ClassABCSolutionsFigure}(a), the points $\mathcal{G}_1,
\ldots, \mathcal{G}_8$) where one of the loop momenta becomes
infinite. Because the order of the pole is related to the ultraviolet power counting
of underlying integrals in the theory (taking into account fermi--bose
cancellations), these insertion points can be
associated, roughly speaking, with UV divergences.

The pattern of global poles in classes (a)--(c) can be understood as
follows.  Starting from fig.~\ref{ClassABCSolutionsFigure}(a), we can
imagine taking massless limits of external
momenta, at each step having exactly one additional double-box rung
end on a three-point vertex.  Geometrically, each step adds a pair of
pinches (nodal points)---the first step producing
fig.~\ref{ClassABCSolutionsFigure}(b), and the second producing
fig.~\ref{ClassABCSolutionsFigure}(c).  Each pinch preserves the
global poles already present, and adds a global pole at the location
of the pinching point.  Nonetheless, pinching leaves the number of
\emph{independent} global poles constant: while it creates a new
global pole, it also creates a new Riemann sphere and hence adds a
global residue constraint which allows one global pole to be
eliminated.  There are eight independent global poles in class~(a),
and the number remains eight in classes~(b) and (c).

To be more specific, class~(b) contains 12 global poles, as illustrated in
fig.~\ref{ClassABCSolutionsFigure}(b).  The poles
$\mathcal{G}_1,\ldots,\mathcal{G}_{10}$ are obtained by taking the limit
$\xib'_3 \to 0$ (corresponding to either $m_3 \to 0$ or $m_4 \to 0$) of the
class~(a) poles in eq.~(\ref{ClassGlobalPoles}).\footnote{For convenience,
  we note that the labeling of global poles here is related to that of
  ref.~\cite{ExternalMasses} as follows:
\begin{align}
   \big(\mathcal{G}_1^\mathrm{(b)}, \mathcal{G}_2^\mathrm{(b)},
        \mathcal{G}_3^\mathrm{(b)}, \mathcal{G}_4^\mathrm{(b)},
   \mathcal{G}_5^\mathrm{(b)}, \mathcal{G}_6^\mathrm{(b)},
   \mathcal{G}_7^\mathrm{(b)}, \mathcal{G}_8^\mathrm{(b)}
   \big)^\mathrm{\text{\scriptsize \cite{ExternalMasses}}}
   \hspace{1mm}&=\hspace{1mm} \big(\mathcal{G}_9^\mathrm{(b)},
    \mathcal{G}_{10}^\mathrm{(b)}, \mathcal{G}_{11}^\mathrm{(b)},
    \mathcal{G}_{12}^\mathrm{(b)},
   \mathcal{G}_3^\mathrm{(b)}, \mathcal{G}_4^\mathrm{(b)},
   \mathcal{G}_5^\mathrm{(b)}, \mathcal{G}_6^\mathrm{(b)} \big) \nn\\
   \big(\mathcal{G}_1^\mathrm{(c)}, \mathcal{G}_2^\mathrm{(c)},
        \mathcal{G}_3^\mathrm{(c)}, \mathcal{G}_4^\mathrm{(c)},
   \mathcal{G}_5^\mathrm{(c)}, \mathcal{G}_6^\mathrm{(c)},
     \mathcal{G}_7^\mathrm{(c)}, \mathcal{G}_8^\mathrm{(c)}
   \big)^\mathrm{\text{\scriptsize\cite{ExternalMasses}}}
   \hspace{1mm}&=\hspace{1mm} \big(\mathcal{G}_9^\mathrm{(c)}, \mathcal{G}_{10}^\mathrm{(c)}, \mathcal{G}_{11}^\mathrm{(c)}, \mathcal{G}_{12}^\mathrm{(c)},
   \mathcal{G}_{14}^\mathrm{(c)}, \mathcal{G}_{13}^\mathrm{(c)}, \mathcal{G}_5^\mathrm{(c)}, \mathcal{G}_6^\mathrm{(c)} \big)\nn
\end{align}
where the labeling on the left-hand sides corresponds to that of
eqs.~(4.16) and (4.27) of ref.~\cite{ExternalMasses}. The
labeling on the right-hand sides is that of the present paper,
with the superscript ${}^\mathrm{(b)}$ denoting that one should take the
$\xib'_3 \to 0$ limit of the poles listed in eq.~(\ref{ClassGlobalPoles})
to obtain the class~(b) poles, and the superscript ${}^\mathrm{(c)}$
indicating that one should further take the limit $\xib'_2 \to 0$ to
find the class~(c) poles.}

As evaluating the limit of $\mathcal{G}_9$
and $\mathcal{G}_{10}$ is slightly subtle, we quote the result here:
\begin{equation}
\begin{aligned}
\Global_9:
\lim_{\xib'_3 \to 0} \bigg(z_+, \hspace{0.9mm} \frac{\xib'_2}{z_+}, \hspace{0.9mm} -\frac{(z_+ \xib'_1 + \tau \xib'_2) \xib'_3}
{(z_+ \xib'_1 + \xib'_2) \xib'_4}, \hspace{0.9mm} -\frac{(z_+ \xib'_1 + \xib'_2) \xib'_4}
{z_+ \xib'_1 + \tau \xib'_2} \bigg) &= \bigg(-\frac{\xib'_2}{\xib'_1},-\xib'_1,\frac{\xib'_2-\xib'_1}{\xib'_1-\tau  \xib'_2},0\bigg)\,, \\
\Global_{10}:
\lim_{\xib'_3 \to 0} \bigg(z_-, \hspace{0.9mm} \frac{\xib'_2}{z_-}, \hspace{0.9mm} -\frac{1+z_-}{1+\tau z_-}, \hspace{0.9mm}
-\frac{(1+\tau z_-)\xib'_3}{1+z_-}\bigg) &= \bigg(-1,-\xib'_2,0,-\frac{\left(\xib'_1-\xib'_2\right) \xib'_4}{\xib'_1-\tau \xib'_2}\bigg) \,.
\end{aligned}
\label{eq:ClassB_Limit_of_Global_Poles}
\end{equation}
In addition, class~(b) contains the following two global poles
\begin{equation}
\begin{aligned}
\Global_{11}:~
&\bigg(-\frac{\xib'_2}{\xib'_1},-\xib'_1,0,0\bigg)\,,\\
\Global_{12}:~
\;&\big(-1,-\xib'_2,0,0\big) \,,
\end{aligned}
\end{equation}
which are exactly the two nodal points created during the pinches
$\mathcal{S}_1 \to \mathcal{S}_1 \cup \mathcal{S}_4$ and $\mathcal{S}_2 \to
\mathcal{S}_2 \cup \mathcal{S}_3$ (compare
figs.~\ref{ClassABCSolutionsFigure}(a) and
\ref{ClassABCSolutionsFigure}(b)).  Similarly, class~(c) contains 14 global
poles, as illustrated in fig.~\ref{ClassABCSolutionsFigure}(c).  The poles
$\mathcal{G}_1,\ldots,\mathcal{G}_{12}$ are obtained by taking the limit
$\xib'_2 \to 0$ (corresponding to either $m_1 \to 0$ or $m_2 \to 0$) of the
class~(b) global poles. In addition, class~(c) contains the following two
global poles,
\begin{equation}
\begin{aligned}
\Global_{13}:~
&\big(0,0,0,-\xib'_4\big)\,,\\
\Global_{14}:~
&\big(0,0,-1,0\big) \,,
\end{aligned}
\end{equation}
which are precisely the two nodal points created during the pinches $\mathcal{S}_4 \to \mathcal{S}_4 \cup \mathcal{S}_6$
and $\mathcal{S}_2 \to \mathcal{S}_2 \cup \mathcal{S}_5$ (compare figs.~\ref{ClassABCSolutionsFigure}(b) and \ref{ClassABCSolutionsFigure}(c)).

As explained in sec.~\ref{sec:four-mass_projectors}, contours for GDOs
must annihilate any function that integrates to zero on $\mathbb{R}^D
\times \mathbb{R}^D$.  In terms of the coefficients $\omega_i$ of the
basis contours $C_j$, this requirement yields the following
constraints from numerator insertions of Levi-Civita tensors (which are
parity odd):
\begin{equation}
\begin{array}{rll}
2\omega_1 - 2\omega_2 - \omega_9 + \omega_{10} + \omega_{11}
  - \omega_{12} + \omega_{13} - \omega_{14}  &=&  0 \,\phantom{.}\\[1mm]
2\omega_3 - 2\omega_4 - \omega_9 + \omega_{10} - \omega_{11}
  + \omega_{12} + \omega_{13} - \omega_{14}  &=&  0 \,\phantom{.}\\[1mm]
2\omega_5 - 2\omega_6 - \omega_9 + \omega_{10} - \omega_{11}
  + \omega_{12} + \omega_{13} - \omega_{14}  &=&  0 \,\phantom{.}\\[1mm]
2\omega_7 - 2\omega_8 - \omega_9 + \omega_{10} - \omega_{11}
  + \omega_{12} - \omega_{13} + \omega_{14}  &=&  0 \,,
\end{array}\label{eq:ClassC_LeviCivita_constraints}
\end{equation}
where $\omega_{11,12,13,14}\to 0$ in class~(a), and $\omega_{13,14}\to 0$
in class~(b).  If we choose a homology basis consisting of parity-conjugate
pairs of poles, these constraints are expressed by the simple geometric
statement that a valid contour must be invariant under a rotation through
$\pi$ radians of figs.~\ref{ClassABCSolutionsFigure}(a),
\ref{ClassABCSolutionsFigure}(b), and \ref{ClassABCSolutionsFigure}(c),
respectively.

Consistency with IBP identities imposes less
transparent constraints on maximal-cut contours.
In class (b), there is a single IBP constraint which takes the form,
\begin{equation}
2\omega_1 + 2\omega_2 - \omega_3 - \omega_4 - \omega_7 - \omega_8 - \omega_9 - \omega_{10} + \omega_{11} + \omega_{12} = 0 \,,
\label{eq:ClassB_IBP_constraint}
\end{equation}
whereas in class~(c) there are two IBP constraints which take the form,
\begin{align}
2\omega_1 + 2\omega_2 -  \omega_3 -  \omega_4 -  \omega_7 -  \omega_8
- \omega_9 - \omega_{10} + \omega_{11} + \omega_{12}                     &=  0\,,
\label{eq:ClassC_IBP_constraint_1}  \\
 \omega_3 +  \omega_4 + 2\omega_5 + 2\omega_6 - 3\omega_7 - 3\omega_8
- \omega_9 - \omega_{10} - \omega_{11} - \omega_{12}
+ 2\omega_{13} + 2\omega_{14}  &=  0
\label{eq:ClassC_IBP_constraint_2} \,.
\end{align}
The first class~(c)
constraint~(\ref{eq:ClassC_IBP_constraint_1}) is identical to the
class~(b) one in eq.~(\ref{eq:ClassB_IBP_constraint}).  This suggests
that these constraints arise during the pinchings that carry the
doubly pinched torus depicted in fig.~\ref{ClassABCSolutionsFigure}(a)
into the quadruply pinched torus of
fig.~\ref{ClassABCSolutionsFigure}(b) and thence into the sextuply
pinched torus of fig.~\ref{ClassABCSolutionsFigure}(c). The
transition from fig.~\ref{ClassABCSolutionsFigure}(a) into
fig.~\ref{ClassABCSolutionsFigure}(b) involves two (parity-conjugate)
pinches which one might at first expect to produce two
constraints. However, as a valid contour must be
parity-symmetric~(\ref{eq:ClassC_LeviCivita_constraints}), we should
really expect \emph{one} independent constraint to arise from a
double pinching.  This constraint is accompanied by a second
constraint arising from the double pinching that turns
fig.~\ref{ClassABCSolutionsFigure}(b) into
fig.~\ref{ClassABCSolutionsFigure}(c). This pattern offers hope that
it may be possible to derive
the IBP
constraints~(\ref{eq:ClassB_IBP_constraint})--(\ref{eq:ClassC_IBP_constraint_2})
directly from the underlying algebraic geometry.

Expressing the IBP constraints in an overcomplete basis of homology
makes it clear that they cannot be determined from algebraic
\emph{topology} alone. For example, on the sphere $\mathcal{S}_4$ in
fig.~\ref{ClassABCSolutionsFigure}(b), the poles $\mathcal{G}_3,
\mathcal{G}_5, \mathcal{G}_7$ may be freely relabeled among each other
without changing the topology. In contrast,
eq.~(\ref{eq:ClassB_IBP_constraint}) does not have this relabeling
symmetry.

\subsection{Discrete Symmetries of IBP Constraints}

We observe that the class~(b) IBP constraint (\ref{eq:ClassB_IBP_constraint})
is symmetric under reflection of fig.~\ref{ClassABCSolutionsFigure}(b) in the vertical axis
passing through the poles $\mathcal{G}_1$ and $\mathcal{G}_2$. More explicitly,
eq.~(\ref{eq:ClassB_IBP_constraint}) is symmetric under the
interchanges\footnote{This symmetry does not have an obvious physical meaning:
it corresponds to flipping the right loop of the double-box
graph through a vertical axis.}
\begin{equation}
\begin{array}{rllrll}
\omega_1 \hspace{-2mm}&\longleftrightarrow&\hspace{-1.5mm} \omega_1  & \hspace{14mm} \omega_5    \hspace{-2mm}&\longleftrightarrow&\hspace{-1.5mm}  \omega_6  \\
\omega_2 \hspace{-2mm}&\longleftrightarrow&\hspace{-1.5mm} \omega_2  & \hspace{14mm} \omega_9    \hspace{-2mm}&\longleftrightarrow&\hspace{-1.5mm} -\omega_{11}  \\
\omega_3 \hspace{-2mm}&\longleftrightarrow&\hspace{-1.5mm} \omega_8  & \hspace{14mm} \omega_{10} \hspace{-2mm}&\longleftrightarrow&\hspace{-1.5mm} -\omega_{12} \,.  \\
\omega_4 \hspace{-2mm}&\longleftrightarrow&\hspace{-1.5mm} \omega_7  &
\end{array} \label{eq:symmetry_of_ClassB_IBP_constraint}
\end{equation}
The pattern of relative minuses in eq.~(\ref{eq:symmetry_of_ClassB_IBP_constraint})
owes to the fact that, in our orientation conventions, the reflection flips
the orientation of the pinching or `IR' cycles, but preserves that of the insertion or `UV' cycles.

Conversely, assuming the symmetry (\ref{eq:symmetry_of_ClassB_IBP_constraint}), one might ask to what
extent it determines the IBP constraint. The most general IBP constraint
invariant under eq.~(\ref{eq:symmetry_of_ClassB_IBP_constraint}) takes the form
\begin{equation}
a_1 \omega_1  +  a_2 \omega_2  +  a_3 (\omega_3 + \omega_8)  +  a_4 (\omega_4 + \omega_7)
+  a_5 (\omega_5 + \omega_6)  +  a_6 (\omega_9 - \omega_{11}) +  a_7 (\omega_{10} - \omega_{12}) = 0 \,.
\label{eq:ClassB_IBP_constraint_disc_symm_1}
\end{equation}
For convenience, let us now choose a basis of homology, for example
$\omega_{1,2,5,6}=0$. In this basis, the IBP constraint
(\ref{eq:ClassB_IBP_constraint_disc_symm_1}) takes the form,
\begin{equation}
r^\mathrm{(b)}_1 \big(\omega_3 + \omega_4 + \omega_7 + \omega_8 \big)
+ r^\mathrm{(b)}_2 \big(\omega_9 + \omega_{10} - \omega_{11} - \omega_{12} \big) \hspace{0.5mm}=\hspace{0.5mm} 0 \,,
\end{equation}
where we furthermore imposed the Levi-Civita constraints~(\ref{eq:ClassC_LeviCivita_constraints}).
Remarkably, the only thing left unexplained by the flip symmetry (\ref{eq:symmetry_of_ClassB_IBP_constraint})
is the fact that $r^\mathrm{(b)}_1 = r^\mathrm{(b)}_2 \neq 0$.

Out of the two IBP constraints in class~(c), we observe that eq.~(\ref{eq:ClassC_IBP_constraint_1})
is inherited directly from eq.~(\ref{eq:ClassB_IBP_constraint}) whereas
the difference between eq.~(\ref{eq:ClassC_IBP_constraint_1}) and eq.~(\ref{eq:ClassC_IBP_constraint_2})
is symmetric under reflection of fig.~\ref{ClassABCSolutionsFigure}(c) in a line
passing through the centers of the spheres $\mathcal{S}_5$ and $\mathcal{S}_6$. More explicitly,
the difference is symmetric under the interchanges
\begin{equation}
\begin{array}{rllrll}
\omega_1 \hspace{-2mm}&\longleftrightarrow&\hspace{-1.5mm} \omega_7  & \hspace{14mm} \omega_9    \hspace{-2mm}&\longleftrightarrow&\hspace{-1.5mm} -\omega_{10}  \\
\omega_2 \hspace{-2mm}&\longleftrightarrow&\hspace{-1.5mm} \omega_8  & \hspace{14mm} \omega_{11} \hspace{-2mm}&\longleftrightarrow&\hspace{-1.5mm} -\omega_{13}  \\
\omega_3 \hspace{-2mm}&\longleftrightarrow&\hspace{-1.5mm} \omega_5  & \hspace{14mm} \omega_{12} \hspace{-2mm}&\longleftrightarrow&\hspace{-1.5mm} -\omega_{14} \,.  \\
\omega_4 \hspace{-2mm}&\longleftrightarrow&\hspace{-1.5mm} \omega_6  &
\end{array} \label{eq:symmetry_of_ClassC_IBP_constraint}
\end{equation}
This symmetry will be broken by any choice of homology
basis, highlighting the virtue of expressing the IBP constraints
(\ref{eq:ClassC_IBP_constraint_1})--(\ref{eq:ClassC_IBP_constraint_2}) in
an overcomplete basis.

In analogy with the above, we can write down the most general
constraint invariant under eq.~(\ref{eq:symmetry_of_ClassC_IBP_constraint}),
choose a basis of homology such as $\omega_{1,2,5,6,7,8}=0$
and impose the Levi-Civita constraints~(\ref{eq:ClassC_LeviCivita_constraints}). We are
then left with the constraint,
\begin{equation}
r^\mathrm{(c)}_1 \big(\omega_3 + \omega_4 \big)
+ r^\mathrm{(c)}_2 \big(\omega_{11} + \omega_{12} - \omega_{13} - \omega_{14} \big) \hspace{0.5mm}=\hspace{0.5mm} 0 \,.
\end{equation}
Only the requirement that $r^\mathrm{(c)}_1 = -r^\mathrm{(c)}_2 \neq 0$ is
left unexplained by the flip symmetry
(\ref{eq:symmetry_of_ClassC_IBP_constraint}).

\section{Conclusions}
\label{sec:Conclusions}

In this paper we have extended the maximal-unitarity formalism at two loops
to double-box integrals with four massive external legs.  We have constructed
generalized discontinuity operators which isolate each of the four
master integrals, annihilating all others.  Applying one of these GDOs to
the amplitude yields a formula for the corresponding coefficient in
\eqn{BasicEquation}, as a contour integral over products of tree-level
amplitudes.

We can choose to think of each GDO as operating in two steps.  In the first
step, we perform seven of the eight contour integrals, thereby putting on
shell all internal lines of the double-box integral.  This
restricts the integrand to a Riemann surface, which has the form of a
multiply-pinched torus.  Each component is a Riemann sphere, with the
number of spheres equal to twice the number of double-box rungs that end on
a three-point vertex.  This step is identical for all four GDOs.

In the second step, we perform the remaining contour integral over a
contour on the Riemann surface.  This fully localizes the integrand onto a
combination of global poles.  The integration contours are different for
each GDO.  They are subject to consistency constraints.  These constraints
fall into two classes for general double-box integrals: (a) parity
symmetry, amounting to invariance of the contours under rotations through
$\pi$ radians of the pinched tori; and (b) consistency with IBP relations.
Writing out the latter constraints in an overcomplete basis of homology
exposes additional flip symmetries.  These symmetries alone would allow us to
determine the constraints up to a small number of constants.  The IBP
relations determine these constants.

For the four-mass double box (class (a)), the underlying Riemann surface
consists of two spheres, and there is no contour constraint from IBP
relations.  For the three-mass and short-side two-mass double boxes (class
(b)), considered previously in ref.~\cite{ExternalMasses}, the Riemann
surface consists of four linked spheres.  It can be viewed as derived
from the two-sphere Riemann surface via a double pinching.  One IBP
constraint arises here.  This constraint is inherited by the last case, a
six-sphere surface corresponding to massless, one-mass, diagonal and
long-side two-mass double boxes (class (c)), considered previously in
refs.~\cite{MaximalTwoLoopUnitarity,ExternalMasses}.  The six spheres again
can be viewed as derived from the four-sphere surface via a double
pinching, and an additional IBP constraint emerges as well.  Thus, the IBP
contour constraints appear to arise during the chiral branchings of the
on-shell solutions, suggesting a strong connection to the underlying
algebraic geometry.

This sequence of IBP constraints suggests a more natural choice of master
integrals than that of \eqn{eq:master_integrals}. Namely, one can construct
a set of four integrals with the property that in class~(a) all integrals
are linearly independent, whereas in classes~(b) and (c) respectively
one and two elements become zero (up to terms with vanishing heptacuts),
by virtue of integration-by-parts relations. (We refer to
App.~\ref{sec:IBP-inspired_basis} for an explicit construction
of such a set of integrals.)

A complete calculation of four-point amplitudes will also require
the $\Ord(\epsilon)$ terms in integral coefficients, and also
GDOs for integrals with fewer than seven propagators. For processes
with additional external legs, higher-point integrals will be needed
as well.  The simplest extension would probably be to `turtle-box'
integrals ($P^*_{2,2}$ in the notation of ref.~\cite{TwoLoopBasis}),
as their properties are related to those of the double-box integrals
considered here and in refs.~\cite{MaximalTwoLoopUnitarity,ExternalMasses}.

The generalized discontinuity operators whose contours are given by
eqs.~(\ref{GeneralContour}) and~(\ref{eq:master_contour}), along with similar
results from ref.~\cite{ExternalMasses}, can be applied directly to
computations of two-loop amplitudes in both numerical and analytic forms.
Their construction also hints at deeper connections to the algebraic
geometry of the corresponding Feynman integrals.

\section*{Acknowledgments}

K.~J.~L. thanks CERN, where part of this work was carried out, for its hospitality. We also
thank Spencer Bloch, Simon Caron-Huot, Benjamin Matschke and Masahiko
Taniguchi for useful discussions. This work was supported by the Research
Executive Agency (REA) of the European Union under the Grant Agreement
number PITN--GA--2010--264564 (LHCPhenoNet).  This work is supported by the
European Research Council under Advanced Investigator Grant
ERC--AdG--228301.

\appendix

\section{Explicit Loop Momenta}
\label{ExplicitMomentaAppendix}

In this appendix, we present explicit forms for the loop momenta in the two
solutions, ${\cal S}_{1,2}$, of the four-mass planar double-box heptacut
equations.  Solution ${\cal S}_2$ is given explicitly by
\begin{equation}
\begin{aligned}
\ell_1^\mu&=
\xib'_1  \Big(k_1^{\flat,\mu}
+ \frac{z}{2} \frac{\spa{1^\flat}.{4^\flat}}{\spa{2^\flat}.{4^\flat}}
       \ssand{2^\flat}.{\gamma^\mu}.{1^\flat}\Big)
+ \xib'_2 \frac{\projdot1.4}{\projdot2.4} \Big(k_2^{\flat,\mu}
 + \frac{1}{2z} \frac{\spa{2^\flat}.{4^\flat}}{\spa{1^\flat}.{4^\flat}} \ssand{1^\flat}.{\gamma^\mu}.{2^\flat}\Big)\,, \\
 \ell_2^\mu&=
\xib'_4 \Big( k_4^{\flat,\mu}  + \frac{w}{2} \frac{\spa{1^\flat}.{4^\flat}}{\spa{1^\flat}.{3^\flat}}  \ssand{3^\flat}.{\gamma^\mu}.{4^\flat}\Big) +
\xib'_3 \frac{\projdot1.4}{\projdot1.3} \Big(k_3^{\flat,\mu}
+  \frac{1}{2w} \frac{\spa{1^\flat}.{3^\flat}}{\spa{1^\flat}.{4^\flat}}  \ssand{4^\flat}.{\gamma^\mu}.{3^\flat}\Big) \,,
\end{aligned}
\end{equation}
where
\begin{equation}
w = -\frac{z+1 }{z \tau+1}\,.
\end{equation}

Solution ${\cal S}_2$ has poles corresponding to infinite momenta
located at $z=\{0,\infty,-1,-1/\tau\}$, or alternatively, at
$w=\{-1,-1/\tau,0,\infty\}$. For the first two $\ell_1^\mu$ is
infinite, and $\ell_2^\mu$ takes the finite values,
\begin{equation}
\hspace{-2mm}\begin{aligned}
&\mathcal{G}_8\!: \ell_2^\mu (z=0)=
\xib'_4 \Big( k_4^{\flat,\mu}  - \frac{1}{2} \frac{\spa{1^\flat}.{4^\flat}}{\spa{1^\flat}.{3^\flat}}  \ssand{3^\flat}.{\gamma^\mu}.{4^\flat}\Big) +
\xib'_3 \frac{\projdot1.4}{\projdot1.3} \Big(k_3^{\flat,\mu}  -  \frac{1}{2}\frac{\spa{1^\flat}.{3^\flat}}{\spa{1^\flat}.{4^\flat}}  \ssand{4^\flat}.{\gamma^\mu}.{3^\flat}\Big)\,,\\
&\mathcal{G}_4\!: \ell_2^\mu (z=\infty)=
\xib'_4 \Big( k_4^{\flat,\mu}  -\frac{1}{2} \frac{\spa{2^\flat}.{4^\flat}}{\spa{2^\flat}.{3^\flat}}  \ssand{3^\flat}.{\gamma^\mu}.{4^\flat}\Big) +
\xib'_3 \frac{\projdot1.4}{\projdot1.3} \Big(k_3^{\flat,\mu}  -\frac{1}{2}\frac{\spa{2^\flat}.{3^\flat}}{\spa{2^\flat}.{4^\flat}}  \ssand{4^\flat}.{\gamma^\mu}.{3^\flat}\Big) \,.
\end{aligned}
\hspace{-3mm}\end{equation}
These two values are related by a swap of legs
$1\leftrightarrow2$. For the latter two poles $\ell_2^\mu$ is
infinite, and $\ell_1^\mu$ takes the finite values,
\begin{equation}
\hspace{-3mm}\begin{aligned}
&\mathcal{G}_2\!: \ell_1^\mu(z=-1)=
\xib'_1  \Big(k_1^{\flat,\mu}  - \frac{1}{2}  \frac{\spa{1^\flat}.{4^\flat}}{\spa{2^\flat}.{4^\flat}}   \ssand{2^\flat}.{\gamma^\mu}.{1^\flat}\Big) +
\xib'_2 \frac{\projdot1.4}{\projdot2.4} \Big(k_2^{\flat,\mu}  - \frac{1}{2} \frac{\spa{2^\flat}.{4^\flat}}{\spa{1^\flat}.{4^\flat}} \ssand{1^\flat}.{\gamma^\mu}.{2^\flat}\Big)\,,\\
&\mathcal{G}_6\!: \ell_1^\mu\Big(z=-\frac{1}{\tau}\Big)=
\xib'_1  \Big(k_1^{\flat,\mu}  - \frac{1}{2} \frac{\spa{1^\flat}.{3^\flat}}{\spa{2^\flat}.{3^\flat}}   \ssand{2^\flat}.{\gamma^\mu}.{1^\flat}\Big) +
\xib'_2 \frac{\projdot1.4}{\projdot2.4} \Big(k_2^{\flat,\mu}  - \frac{1}{2} \frac{\spa{2^\flat}.{3^\flat}}{\spa{1^\flat}.{3^\flat}} \ssand{1^\flat}.{\gamma^\mu}.{2^\flat}\Big) \,,
\end{aligned}
\hspace{-3mm}\end{equation}
which are related by the swap of legs $3\leftrightarrow4$. (Moreover,
as should be clear from the left-right symmetry of the double box,
there is a map $\{1,2,\ell_1,z\} \leftrightarrow \{4,3,\ell_2,w\}$
that relates the above two pairs of poles.)

Solution ${\cal S}_1$ can be obtained from ${\cal S}_2$ by spinor
conjugation $\spa{a}.{b} \leftrightarrow \spb{b}.a$, along with the
reparametrization $z\rightarrow z/\xib'_1$.
 The values of the momenta at poles
$z=\{0,\infty,-\xib'_1,-\xib'_1/\tau\}$ are given by,
\begin{equation}
\begin{alignedat}{2}
&\mathcal{G}_7:~~\ell_{i,{\cal S}_1}^\mu (0) =  (\ell_{i,{\cal S}_2}^\mu  (0))^\dagger\,,~~~~~~
&&\mathcal{G}_3:~~\ell_{i,{\cal S}_1}^\mu (\infty) = (\ell_{i,{\cal S}_2}^\mu  (\infty))^\dagger\,, \\
&\mathcal{G}_1:~~\ell_{i,{\cal S}_1}^\mu (-\xib'_1) = (\ell_{i,{\cal S}_2}^\mu(-1))^\dagger\,,~~~~~~
&&\mathcal{G}_5:~~\ell_{i,{\cal S}_1}^\mu (-\xib'_1/\tau ) = (\ell_{i,{\cal S}_2}^\mu  (-1/\tau))^\dagger
\end{alignedat}
\end{equation}
where ${}^\dagger$ denotes spinor conjugation.

At the Jacobian poles $z=z_\pm$ for ${\cal S}_2$, and $z=\xib'_1 z_\pm$ for
${\cal S}_1$, the momenta are finite, and satisfy the relations
\begin{equation}
\begin{alignedat}{2}
\ell_{1,{\cal S}_1}^\mu(\xib'_1z_\pm)&=
   -\ell_{2,{\cal S}_1}^\mu(\xib'_1z_\pm)\,,\qquad
&\ell_{1,{\cal S}_2}^\mu(z_\pm)&=-\ell_{2,{\cal S}_2}^\mu(z_\pm)\,,  \\
\ell_{i,{\cal S}_1}^\mu(\xib'_1 z_\pm)&=\ell_{i,{\cal S}_2}^\mu(z_\mp)\,,\\
\ell_{i,{\cal S}_2}^\mu(z_+)&=(\ell_{i,{\cal S}_2}^\mu(z_-))^\dagger\,,
&\ell_{i,{\cal S}_1}^\mu(\xib'_1z_+)&=(\ell_{i,{\cal S}_1}^\mu(\xib'_1z_-))^\dagger\,.
\end{alignedat}
\end{equation}
The first two relations follow because the Jacobian pole corresponds to
the middle rung in the double box becoming soft,
$\ell_1+\ell_2=0$. The third relation is a consequence of the fact that the
Jacobian pole is located on the intersection of the two spheres,
${\cal S}_1 \cap {\cal S}_2 $.  The
fourth and fifth identities arise because the two Jacobian poles are
complex conjugates. Because $\ell_1+\ell_2=0$, the two distinct kinematic
solutions are identical to the two quadruple-cut solutions for
a one-loop  four-mass box~\cite{BCFUnitarity}.

\section{IBP-Inspired Choice for the Integral Basis}\label{sec:IBP-inspired_basis}

The sequence of IBP constraints
(\ref{eq:ClassB_IBP_constraint})--(\ref{eq:ClassC_IBP_constraint_2})
in classes~(b) and~(c) suggests natural choices of master
integrals.
We work out the details of such a basis here. We note that one can construct a set of four integrals
with the property that in class~(a) all integrals are linearly
independent, whereas in classes~(b) and (c) respectively one and
two basis elements drop out, because they vanish identically or become reducible via IBPs to simpler
topologies.

To construct such a set of integrals, we start from the observation
that in class~(b) there is a unique IBP constraint, corresponding
to a unique double-box numerator insertion which yields
zero after integration. In the labeling of fig.~\ref{fig:DoubleBox}, case (b) corresponds to the vanishing of
the product $m_3 m_4$, but there is an analogous case (b$'$) where $m_1 m_2=0$.
The unique residues of the IBP constraints in these two kinematic cases of the double-box integral are given by,
\begin{align}
\mathcal{R}^{\mathrm{(b)}}_\mathrm{IBP}
\hspace{0.8mm}&=\hspace{0.8mm}  (2, \hspace{0.3mm} 2,
\hspace{0.3mm} -1, \hspace{0.3mm} -1, \hspace{0.3mm} 0,
\hspace{0.3mm} 0, \hspace{0.3mm} -1, \hspace{0.3mm} -1,
\hspace{0.3mm} -1, \hspace{0.3mm} -1)
\hspace{10mm} \mathrm{when} \hspace{8mm} m_3 m_4 =0\,, \label{eq:residues_of_class_(b)_IBP_1}\\
\mathcal{R}^{\mathrm{(b')}}_\mathrm{IBP}
\hspace{0.8mm}&=\hspace{0.8mm}  (-1, \hspace{0.3mm} -1,
\hspace{0.3mm} 0, \hspace{0.3mm} 0, \hspace{0.3mm} -1,
\hspace{0.3mm} -1, \hspace{0.3mm} 2, \hspace{0.3mm} 2,
\hspace{0.3mm} 1, \hspace{0.3mm} 1)
\hspace{16mm} \mathrm{when} \hspace{8mm} m_1 m_2 =0 \,, \label{eq:residues_of_class_(b)_IBP_2}
\end{align}
where the residues correspond to the global poles $(\mathcal{G}_i)_{i=1,\ldots,10}$.
Here, the list of residues in eq.~(\ref{eq:residues_of_class_(b)_IBP_1})
was read off from the left-hand side of
the IBP constraint (\ref{eq:ClassB_IBP_constraint}).
The class~(b$'$)  IBP constraint, for the case $m_1 m_2 =0$,
can be obtained from the $m_3 m_4 =0$ case by relabeling
the global poles and their residues according to the
left-right flip of fig.~\ref{ClassABCSolutionsFigure}(a)
through a vertical axis intersecting the poles $\mathcal{G}_9$ and $\mathcal{G}_{10}$.
(One must take into account the flip in orientation of the
cycles around the nodal points $\mathcal{G}_{9,10}$, which causes their residues to change sign.)
 Note that the second constraint in class (c), eq.~(\ref{eq:ClassC_IBP_constraint_2}), precisely corresponds to the linear combination
$-\mathcal{R}^{\mathrm{(b)}}_\mathrm{IBP}
-2\mathcal{R}^{\mathrm{(b')}}_\mathrm{IBP}$.

We can now construct a pair of new integrals $I_{3}'$ and $I_{4}'$ whose residues are proportional
respectively to the two IBP residues.~(\ref{eq:residues_of_class_(b)_IBP_1},\,\ref{eq:residues_of_class_(b)_IBP_2}).
  The ans\"atze for the new integrals are,
\begin{equation}
I_3' \hspace{0.6mm}=\hspace{0.6mm} \sum_{j=1}^4 a_j I_j \hspace{9mm} \mathrm{and} \hspace{9mm}
I_4' \hspace{0.6mm}=\hspace{0.6mm} \sum_{j=1}^4 b_j I_j \,,
\end{equation}
with $I_j$ denoting the integrals in eqs.~(\ref{eq:master_integrals}).
We determine the coefficients $a_j, b_j$ by requiring that the residues are proportional,
\begin{equation}
\begin{aligned}
\GRes{i} I_3'  \hspace{0.7mm}&=\hspace{0.7mm}  \sum_{j=1}^4 a_j
\GRes{i} I_j \hspace{1.5mm} \propto \hspace{1.5mm} (2, \hspace{0.3mm} 2,
\hspace{0.3mm} -1, \hspace{0.3mm} -1, \hspace{0.3mm} 0,
\hspace{0.3mm} 0, \hspace{0.3mm} -1, \hspace{0.3mm} -1,
\hspace{0.3mm} -1, \hspace{0.3mm} -1)\,, \\
\GRes{i} I_4'  \hspace{0.7mm}&=\hspace{0.7mm}  \sum_{j=1}^4 b_j
\GRes{i} I_j \hspace{1.5mm}\propto \hspace{1.5mm} (-1, \hspace{0.3mm} -1,
\hspace{0.3mm} 0, \hspace{0.3mm} 0, \hspace{0.3mm} -1,
\hspace{0.3mm} -1, \hspace{0.3mm} 2, \hspace{0.3mm} 2,
\hspace{0.3mm} 1, \hspace{0.3mm} 1) \,,
\end{aligned}
\end{equation}
where the residues
$\GResT{i} I_j$ are
given in eqs.~(\ref{eq:master_integrals_residues},\,\ref{eq:definition_of_r_i}).

Solving for $a_j, b_j$, one finds the following basis of integrals
with the desired properties,
\begin{equation}
\begin{aligned}
 I_1'  &=  I_1\,, \\
 I_2'  &=  I_2\,, \\
 I_3'  &=\left( \frac{r_2 (r_7 - r_5)}{r_4} - \frac{2 r_1 r_8}{r_3}-2 r_1 r_2 + r_5+ r_7\right) I_1+  \frac{ r_5-r_7 }{r_4} I_2 + \frac{2 r_8}{r_3} I_3  +2 I_4 \,, \\
 I_4'  &=   \left(\frac{r_1 (r_8 - r_6)}{r_3}-\frac{2r_2 r_7}{r_4}  +2 r_1 r_2 + r_6 + r_8\right) I_1 + \frac{2 r_7}{r_4}I_2 + \frac{r_6-r_8}{r_3} I_3-2 I_4\,,
\end{aligned}
\end{equation}
where the global residues $r_i$ are defined in eq.~(\ref{eq:definition_of_r_i}).
These integrals are linearly independent in class~(a),
whereas in class~(b),  the heptacut of $I_3'$ vanishes for $m_3 m_4 =0$, and
the heptacut of $I_4'$ vanishes for $m_1 m_2 =0$. In class~(c), both $I_3'$ and $I_4'$ have vanishing heptacuts
because in this class $m_1 m_2$ and $m_3 m_4$ vanish simultaneously. This
is consistent with the class (c) IBP constraints in
eqs.~(\ref{eq:ClassC_IBP_constraint_1})--(\ref{eq:ClassC_IBP_constraint_2})
being linear combinations of the constraints corresponding to the residues
in eqs.~(\ref{eq:residues_of_class_(b)_IBP_1})--(\ref{eq:residues_of_class_(b)_IBP_2}).


\begin{thebibliography}{99}

\bibitem{AtlasHiggs}
G.~Aad {\it et al.}  [ATLAS Collaboration],
  Phys.\ Lett.\ B {\bf 716}, 1 (2012)
  [1207.7214 [hep-ex]].

\bibitem{CMSHiggs}
S.~Chatrchyan {\it et al.}  [CMS Collaboration],
Phys.\ Lett.\ B {\bf 716}, 30 (2012)
[1207.7235 [hep-ex]].

\bibitem{EMZW3j}
R.~K.~Ellis, K.~Melnikov and G.~Zanderighi,
JHEP {\bf 0904}, 077 (2009)
[0901.4101 [hep-ph]];
Phys.\ Rev.\  D {\bf 80}, 094002 (2009)
[0906.1445 [hep-ph]];\\
%
K.~Melnikov and G.~Zanderighi,
Phys.\ Rev.\  D {\bf 81}, 074025 (2010)
[0910.3671 [hep-ph]].

\bibitem{BlackHatW3j}
C.~F.~Berger, Z.~Bern, L.~J.~Dixon, F.~Febres Cordero, D.~Forde, T.~Gleisberg,
H.~Ita, D.~A.~Kosower and D.~Ma\^{\i}tre,
Phys.\ Rev.\ Lett.\  {\bf 102}, 222001 (2009)
[0902.2760 [hep-ph]];\\
C.~F.~Berger, Z.~Bern, L.~J.~Dixon, F.~Febres Cordero, D.~Forde,
T.~Gleisberg, H.~Ita, D.~A.~Kosower and D.~Ma\^{\i}tre,
Phys.\ Rev.\  D {\bf 80}, 074036 (2009)
[0907.1984 [hep-ph]].

\bibitem{BlackHatZ3j}
C.~F.~Berger, Z.~Bern, L.~J.~Dixon, F.~Febres~Cordero, D.~Forde,
T.~Gleisberg, H.~Ita, D.~A.~Kosower and D.~Ma\^{\i}tre,
Phys.\ Rev.\ D {\bf 82}, 074002 (2010)
[1004.1659 [hep-ph]].

\bibitem{BlackHatW4j}
C.~F.~Berger, Z.~Bern, L.~J.~Dixon, F.~Febres~Cordero, D.~Forde,
T.~Gleisberg, H.~Ita, D.~A.~Kosower and D.~Ma\^{\i}tre,
Phys.\ Rev.\ Lett.\  {\bf 106}, 092001 (2011)
[1009.2338 [hep-ph]].

\bibitem{BlackHatZ4j}
H.~Ita, Z.~Bern, L.~J.~Dixon, F.~Febres Cordero, D.~A.~Kosower
and D.~Ma\^{\i}tre,
Phys.\ Rev.\ D {\bf 85}, 031501 (2012)
[1108.2229 [hep-ph]].

\bibitem{Badger:2012pf}
S.~Badger, B.~Biedermann, P.~Uwer and V.~Yundin,
Phys.\ Lett.\ B {\bf 718}, 965 (2013)  [1209.0098 [hep-ph]].

\bibitem{BlackHatW5j}
Z.~Bern, L.~J.~Dixon, F.~Febres Cordero, S.~Hoeche, H.~Ita, D.~A.~Kosower, D.~Maitre and K.~J.~Ozeren,
Phys.\ Rev.\ D {\bf 88}, 014025 (2013)  [1304.1253 [hep-ph]].

\bibitem{NNLOThreeJet}
 A.~Gehrmann-De Ridder, T.~Gehrmann, E.~W.~N.~Glover and G.~Heinrich,
  JHEP {\bf 0711}, 058 (2007)
  [0710.0346 [hep-ph]];\\
  S.~Weinzierl,
  Phys.\ Rev.\ Lett.\  {\bf 101}, 162001 (2008)
  [0807.3241 [hep-ph]].

\bibitem{ThreeJetAlphaS}
  G.~Dissertori, A.~Gehrmann-De Ridder, T.~Gehrmann, E.~W.~N.~Glover, G.~Heinrich, G.~Luisoni and H.~Stenzel,
  JHEP {\bf 0908}, 036 (2009)
  [0906.3436 [hep-ph]];\\
  G.~Dissertori, A.~Gehrmann-De Ridder, T.~Gehrmann, E.~W.~N.~Glover, G.~Heinrich and H.~Stenzel,
  Phys.\ Rev.\ Lett.\  {\bf 104}, 072002 (2010)
  [0910.4283 [hep-ph]].

\bibitem{UnitarityMethod}
Z.~Bern, L.~J.~Dixon, D.~C.~Dunbar and D.~A.~Kosower,
Nucl.\ Phys.\ B {\bf 425}, 217 (1994)
[hep-ph/9403226];
%
Nucl.\ Phys.\ B {\bf 435}, 59 (1995)
[hep-ph/9409265];\\
%
Z.\ Bern, L.\ J.\ Dixon and D.\ A.\ Kosower,
Ann.\ Rev.\ Nucl.\ Part.\ Sci.\  {\bf 46}, 109 (1996)
[hep-ph/9602280].

\bibitem{Bern:1995db}
Z.~Bern and A.~G.~Morgan,
Nucl.\ Phys.\  B {\bf 467}, 479 (1996)
[hep-ph/9511336].

\bibitem{Zqqgg}
Z.~Bern, L.~J.~Dixon and D.~A.~Kosower,
Nucl.\ Phys.\  B {\bf 513}, 3 (1998)
[hep-ph/9708239].

\bibitem{DdimensionalI}
Z.~Bern, L.~J.~Dixon, D.~C.~Dunbar and D.~A.~Kosower,
Phys.\ Lett.\  B {\bf 394}, 105 (1997)
[hep-th/9611127].

\bibitem{BCFUnitarity}
R.~Britto, F.~Cachazo and B.~Feng,
Nucl.\ Phys.\  B {\bf 725}, 275 (2005)
[hep-th/0412103].

\bibitem{OtherUnitarity}
R.~Britto, F.~Cachazo and B.~Feng,
Phys.\ Rev.\  D {\bf 71}, 025012 (2005)
[hep-th/0410179];\\
%
S.~J.~Bidder, N.~E.~J.~Bjerrum-Bohr, L.~J.~Dixon and D.~C.~Dunbar,
Phys.\ Lett.\  B {\bf 606}, 189 (2005)
[hep-th/0410296];\\
%
S.~J.~Bidder, N.~E.~J.~Bjerrum-Bohr, D.~C.~Dunbar and W.~B.~Perkins,
Phys.\ Lett.\  B {\bf 612}, 75 (2005)
[hep-th/0502028];\\
%
S.~J.~Bidder, D.~C.~Dunbar and W.~B.~Perkins,
JHEP {\bf 0508}, 055 (2005)
[hep-th/0505249];\\
%
Z.~Bern, N.~E.~J.~Bjerrum-Bohr, D.~C.~Dunbar and H.~Ita,
JHEP {\bf 0511}, 027 (2005)
[hep-ph/0507019];\\
%
N.~E.~J.~Bjerrum-Bohr, D.~C.~Dunbar and W.~B.~Perkins,
JHEP {\bf 0804}, 038 (2008)
[0709.2086 [hep-ph]].

\bibitem{Bootstrap}
Z.~Bern, L.~J.~Dixon and D.~A.~Kosower,
Phys.\ Rev.\  D {\bf 73}, 065013 (2006)
[hep-ph/0507005].

\bibitem{BCFCutConstructible}
R.~Britto, E.~Buchbinder, F.~Cachazo and B.~Feng,
Phys.\ Rev.\  D {\bf 72}, 065012 (2005)
hep-ph/0503132];\\
%
R.~Britto, B.~Feng and P.~Mastrolia,
Phys.\ Rev.\  D {\bf 73}, 105004 (2006)
[hep-ph/0602178];\\
%
P.~Mastrolia,
Phys.\ Lett.\  B {\bf 644}, 272 (2007)
[hep-th/0611091].

\bibitem{BMST}
A.~Brandhuber, S.~McNamara, B.~J.~Spence and G.~Travaglini,
JHEP {\bf 0510}, 011 (2005)
[hep-th/0506068].

\bibitem{OPP}
G.~Ossola, C.~G.~Papadopoulos and R.~Pittau,
Nucl.\ Phys.\  B {\bf 763}, 147 (2007)
[hep-ph/0609007].

\bibitem{OnShellReview}
Z.~Bern, L.~J.~Dixon and D.~A.~Kosower,
Annals Phys.\  {\bf 322}, 1587 (2007)
[0704.2798 [hep-ph]].

\bibitem{Forde}
D.~Forde,
Phys.\ Rev.\  D {\bf 75}, 125019 (2007)
[0704.1835 [hep-ph]].
%

\bibitem{Badger}
S.~D.~Badger,
JHEP {\bf 0901}, 049 (2009)
[0806.4600 [hep-ph]].

\bibitem{DdimensionalII}
C.~Anastasiou, R.~Britto, B.~Feng, Z.~Kunszt and P.~Mastrolia,
Phys.\ Lett.\  B {\bf 645}, 213 (2007)
[hep-ph/0609191];
%
JHEP {\bf 0703}, 111 (2007)
[hep-ph/0612277];\\
W.~T.~Giele, Z.~Kunszt and K.~Melnikov,
JHEP {\bf 0804}, 049 (2008)
[0801.2237 [hep-ph]].

\bibitem{BFMassive}
R.~Britto and B.~Feng,
Phys.\ Rev.\  D {\bf 75}, 105006 (2007)
[hep-ph/0612089].
JHEP {\bf 0802}, 095 (2008)
[0711.4284 [hep-ph]];\\
%
R.~Britto, B.~Feng and P.~Mastrolia,
Phys.\ Rev.\  D {\bf 78}, 025031 (2008)
[0803.1989 [hep-ph]];\\
%
R.~Britto, B.~Feng and G.~Yang,
JHEP {\bf 0809}, 089 (2008)
[0803.3147 [hep-ph]].

\bibitem{BergerFordeReview}
C.~F.~Berger and D.~Forde,
Ann.\ Rev.\ Nucl.\ Part.\ Sci.\  {\bf 60}, 181 (2010)
[0912.3534 [hep-ph]].

\bibitem{Bern:2010qa}
Z.~Bern, J.~J.~Carrasco, T.~Dennen, Y.~T.~Huang and H.~Ita,
Phys.\ Rev.\  D {\bf 83}, 085022 (2011)
[1010.0494 [hep-th]].

\bibitem{Elvang:2013cua}
H.~Elvang and Y.~-t.~Huang,
arXiv:1308.1697 [hep-th].

\bibitem{EGK}
R.~K.~Ellis, W.~T.~Giele and Z.~Kunszt,
JHEP {\bf 0803}, 003 (2008)
[0708.2398 [hep-ph]].

\bibitem{BlackHatI}
C.~F.~Berger, Z.~Bern, L.~J.~Dixon, F.~Febres Cordero, D.~Forde, H.~Ita,
D.~A.~Kosower and D.~Ma\^{\i}tre,
Phys.\ Rev.\ D {\bf 78}, 036003 (2008)
[0803.4180 [hep-ph]].

\bibitem{CutTools}
G.~Ossola, C.~G.~Papadopoulos and R.~Pittau,
JHEP {\bf 0803}, 042 (2008)
[0711.3596 [hep-ph]].

\bibitem{MOPP}
P.~Mastrolia, G.~Ossola, C.~G.~Papadopoulos and R.~Pittau,
JHEP {\bf 0806}, 030 (2008)
[0803.3964 [hep-ph]].

\bibitem{Rocket}
W.~T.~Giele and G.~Zanderighi,
JHEP {\bf 0806}, 038 (2008)
[0805.2152 [hep-ph]];\\
R.~K.~Ellis, W.~T.~Giele, Z.~Kunszt, K.~Melnikov and G.~Zanderighi,
JHEP {\bf 0901}, 012 (2009)
[0810.2762 [hep-ph]].

\bibitem{BlackHatII}
C.~F.~Berger, Z.~Bern, L.~J.~Dixon, F.~Febres Cordero, D.~Forde, T.~Gleisberg,
H.~Ita, D.~A.~Kosower and D.~Ma\^{\i}tre,
Phys.\ Rev.\ Lett.\  {\bf 102}, 222001 (2009)
[0902.2760 [hep-ph]].

\bibitem{CutToolsHelac}
G.~Bevilacqua, M.~Czakon, C.~G.~Papadopoulos, R.~Pittau and M.~Worek,
JHEP {\bf 0909}, 109 (2009)
[0907.4723 [hep-ph]].

\bibitem{Samurai}
P.~Mastrolia, G.~Ossola, T.~Reiter and F.~Tramontano,
JHEP {\bf 1008}, 080 (2010)
[1006.0710 [hep-ph]].

\bibitem{WPlus4}
C.~F.~Berger {\it et al.},
Phys.\ Rev.\ Lett.\  {\bf 106}, 092001 (2011)
[1009.2338 [hep-ph]].

\bibitem{NGluon}
S.~Badger, B.~Biedermann and P.~Uwer,
Comput.\ Phys.\ Commun.\  {\bf 182}, 1674 (2011)
[1011.2900 [hep-ph]].

\bibitem{MadLoop}
V.~Hirschi, R.~Frederix, S.~Frixione, M.~V.~Garzelli, F.~Maltoni and R.~Pittau,
JHEP {\bf 1105}, 044 (2011)
[1103.0621 [hep-ph]].

\bibitem{MastroliaOssola}
P.~Mastrolia and G.~Ossola,
JHEP 1111:014 (2011)
[1107.6041 [hep-ph]].

\bibitem{BadgerFZ}
S.~Badger, H.~Frellesvig and Y.~Zhang,
JHEP {\bf 1204}, 055 (2012)
[1202.2019 [hep-ph]];\\
JHEP {\bf 1208}, 065 (2012)  [1207.2976 [hep-ph]].


arXiv:1207.2976 [hep-ph].

\bibitem{Zhang:2012ce}
Y.~Zhang,
JHEP 1209:042 (2012)
[1205.5707 [hep-th]].

\bibitem{Mastrolia:2012an}
P.~Mastrolia, E.~Mirabella, G.~Ossola and T.~Peraro,
Phys.\ Lett.\  B {\bf 718}, 173 (2012)
[1205.7087 [hep-ph]].

\bibitem{Kleiss:2012yv}
R.~H.~P.~Kleiss, I.~Malamos, C.~G.~Papadopoulos and R.~Verheyen,
JHEP 1212:038 (2012)
[1206.4180 [hep-ph]].

\bibitem{Mastrolia:2012wf}
 P.~Mastrolia, E.~Mirabella, G.~Ossola and T.~Peraro,
 Phys.\ Rev.\ D {\bf 87}, 085026 (2013)
 [1209.4319 [hep-ph]].

\bibitem{Mastrolia:2012du}
P.~Mastrolia, E.~Mirabella, G.~Ossola, T.~Peraro and H.~van Deurzen,
PoS LL 2012:028 (2012)
[1209.5678 [hep-ph]].

\bibitem{Huang:2013kh}
R.~Huang and Y.~Zhang,
JHEP 1304:080 (2013)
[1302.1023 [hep-ph]].

\bibitem{Mastrolia:2013kca}
P.~Mastrolia, E.~Mirabella, G.~Ossola and T.~Peraro,
[1307.5832 [hep-ph]].

\bibitem{ArkaniHamed:2009dn}
N.~Arkani-Hamed, F.~Cachazo, C.~Cheung and J.~Kaplan,
JHEP {\bf 1003}, 020 (2010)
[0907.5418 [hep-th]].

\bibitem{ArkaniHamed:2009dg}
N.~Arkani-Hamed, J.~Bourjaily, F.~Cachazo and J.~Trnka,
JHEP {\bf 1101}, 049 (2011)  [0912.4912 [hep-th]].

\bibitem{ArkaniHamed:2010kv}
N.~Arkani-Hamed, J.~L.~Bourjaily, F.~Cachazo, S.~Caron-Huot and J.~Trnka,
JHEP {\bf 1101}, 041 (2011)
[1008.2958 [hep-th]].

\bibitem{ArkaniHamed:2010gg}
N.~Arkani-Hamed, J.~L.~Bourjaily, F.~Cachazo, A.~Hodges and J.~Trnka,
JHEP {\bf 1204}, 081 (2012)  [1012.6030 [hep-th]].

\bibitem{ArkaniHamed:2010gh}
N.~Arkani-Hamed, J.~L.~Bourjaily, F.~Cachazo and J.~Trnka,
JHEP {\bf 1206}, 125 (2012)
[1012.6032 [hep-th]].

\bibitem{ArkaniHamed:2012nw}
N.~Arkani-Hamed, J.~L.~Bourjaily, F.~Cachazo, A.~B.~Goncharov, A.~Postnikov and J.~Trnka,
arXiv:1212.5605 [hep-th].

\bibitem{Feng:2012bm}
B.~Feng and R.~Huang,
JHEP 1302:117 (2013)
[1209.3747 [hep-ph]].

\bibitem{Bern:1997nh}
Z.~Bern, J.~S.~Rozowsky and B.~Yan,
Phys.\ Lett.\  B {\bf 401}, 273 (1997)
[hep-ph/9702424].

\bibitem{ABDK}
  C.~Anastasiou, Z.~Bern, L.~J.~Dixon and D.~A.~Kosower,
  Phys.\ Rev.\ Lett.\  {\bf 91}, 251602 (2003)
  [hep-th/0309040].

\bibitem{Bern:2005iz}
  Z.~Bern, L.~J.~Dixon and V.~A.~Smirnov,
  Phys.\ Rev.\  D {\bf 72}, 085001 (2005)
  [hep-th/0505205].

\bibitem{Bern:2006vw}
  Z.~Bern, M.~Czakon, D.~A.~Kosower, R.~Roiban and V.~A.~Smirnov,
  Phys.\ Rev.\ Lett.\  {\bf 97}, 181601 (2006)
  [hep-th/0604074].

\bibitem{Bern:2006ew}
  Z.~Bern, M.~Czakon, L.~J.~Dixon, D.~A.~Kosower and V.~A.~Smirnov,
  Phys.\ Rev.\  D {\bf 75}, 085010 (2007)
  [hep-th/0610248].

\bibitem{Bern:2008ap}
Z.~Bern, L.~J.~Dixon, D.~A.~Kosower, R.~Roiban, M.~Spradlin, C.~Vergu and A.~Volovich,
Phys.\ Rev.\  D {\bf 78}, 045007 (2008)
[0803.1465 [hep-th]].

\bibitem{Kosower:2010yk}
D.~A.~Kosower, R.~Roiban and C.~Vergu,
Phys.\ Rev.\  D {\bf 83}, 065018 (2011)
[1009.1376 [hep-th]].

\bibitem{Bern:2011rj}
Z.~Bern, C.~Boucher-Veronneau and H.~Johansson,
Phys.\ Rev.\ D {\bf 84}, 105035 (2011)
[1107.1935 [hep-th]].

\bibitem{Bern:2000dn}
Z.~Bern, L.~J.~Dixon and D.~A.~Kosower,
JHEP {\bf 0001}, 027 (2000)
[hep-ph/0001001].

\bibitem{Bern:2002tk}
Z.~Bern, A.~De Freitas and L.~J.~Dixon,
JHEP {\bf 0203}, 018 (2002)
[hep-ph/0201161].

\bibitem{BernDeFreitasDixonTwoPhoton}
Z.~Bern, A.~De Freitas and L.~J.~Dixon,
JHEP {\bf 0109}, 037 (2001)
[hep-ph/0109078].

\bibitem{Bern:2002zk}
Z.~Bern, A.~De Freitas, L.~J.~Dixon and H.~L.~Wong,
Phys.\ Rev.\  D {\bf 66}, 085002 (2002)
[hep-ph/0202271].

\bibitem{Bern:2003ck}
  Z.~Bern, A.~De Freitas and L.~J.~Dixon,
  JHEP {\bf 0306}, 028 (2003)
  [hep-ph/0304168].

\bibitem{TwoLoopSplitting}
  Z.~Bern, L.~J.~Dixon and D.~A.~Kosower,
  JHEP {\bf 0408}, 012 (2004)
  [hep-ph/0404293].

\bibitem{DeFreitas:2004tk}
  A.~De Freitas and Z.~Bern,
  JHEP {\bf 0409}, 039 (2004)
  [hep-ph/0409007].

\bibitem{Bern:2007ct}
  Z.~Bern, J.~J.~M.~Carrasco, H.~Johansson and D.~A.~Kosower,
  Phys.\ Rev.\  D {\bf 76}, 125020 (2007)
  [0705.1864 [hep-th]].

\bibitem{Bern:2008pv}
  Z.~Bern, J.~J.~M.~Carrasco, L.~J.~Dixon, H.~Johansson and R.~Roiban,
  Phys.\ Rev.\  D {\bf 78}, 105019 (2008)
  [0808.4112 [hep-th]].

\bibitem{LeadingSingularity}
F.~Cachazo,
arXiv:0803.1988 [hep-th];\\
F.~Cachazo, M.~Spradlin and A.~Volovich,
Phys.\ Rev.\  D {\bf 78}, 105022 (2008)
[0805.4832 [hep-th]].

\bibitem{Bern:2010tq}
  Z.~Bern, J.~J.~M.~Carrasco, L.~J.~Dixon, H.~Johansson and R.~Roiban,
  Phys.\ Rev.\  D {\bf 82}, 125040 (2010)
  [1008.3327 [hep-th]].

\bibitem{Carrasco:2011hw}
J.~J.~M.~Carrasco and H.~Johansson,
J.\ Phys.\ A {\bf 44}, 454004 (2011)
[1103.3298 [hep-th]].

\bibitem{Carrasco:2011mn}
J.~J.~M.~Carrasco and H.~Johansson,
Phys.\ Rev.\ D {\bf 85}, 025006 (2012)
[1106.4711 [hep-th]].

\bibitem{Bern:2012uc}
Z.~Bern, J.~J.~M.~Carrasco, H.~Johansson and R.~Roiban,
Phys.\ Rev.\ Lett.\  109:241602 (2012)
[1207.6666 [hep-th]].

\bibitem{MaximalTwoLoopUnitarity}
  D.~A.~Kosower and K.~J.~Larsen,
  Phys.\ Rev.\ D {\bf 85}, 045017 (2012)
  [1108.1180 [hep-th]].

\bibitem{ExternalMasses}
H.~Johansson, D.~A.~Kosower and K.~J.~Larsen,
Phys.\ Rev.\ D87:025030 (2013)
[1208.1754 [hep-th]].

\bibitem{Loops&Legs}
H.~Johansson, D.~A.~Kosower and K.~J.~Larsen,
PoS LL 2012:066 (2012)
[1212.2132 [hep-th]].

\bibitem{Caron-HuotLarsen}
S.~Caron-Huot and K.~J.~Larsen,
JHEP {\bf 1210}, 026 (2012)
[1205.0801 [hep-ph]].

\bibitem{TwoLoopBasis}
J.~Gluza, K.~Kajda and D.~A.~Kosower,
Phys.\ Rev.\  D {\bf 83}, 045012 (2011)
[1009.0472 [hep-th]].

\bibitem{Henn:2013pwa}
J.~M.~Henn,
Phys.\ Rev.\ Lett.\  {\bf 110}, 251601 (2013)
[1304.1806 [hep-th]].

\bibitem{Henn:2013tua}
J.~M.~Henn, A.~V.~Smirnov and V.~A.~Smirnov,
JHEP {\bf 1307}, 128 (2013)
[1306.2799 [hep-th]].

\bibitem{Henn:2013woa}
J.~M.~Henn and V.~A.~Smirnov,
arXiv:1307.4083 [hep-th].

\bibitem{Sogaard}
M.~Sogaard,
arXiv:1306.1496 [hep-th].

\bibitem{IBP}
F.~V.~Tkachov,
Phys.\ Lett.\ B {\bf 100}, 65 (1981); \\
K.~G.~Chetyrkin and F.~V.~Tkachov,
Nucl.\ Phys.\ B {\bf 192}, 159 (1981).

\bibitem{Laporta}
S.~Laporta,
  Phys.\ Lett.\  B {\bf 504}, 188 (2001)
  [hep-ph/0102032].
S.~Laporta,
Int.\ J.\ Mod.\ Phys.\ A {\bf 15}, 5087 (2000)
[hep-ph/0102033].

\bibitem{GehrmannRemiddi}
T.~Gehrmann and E.~Remiddi,
Nucl.\ Phys.\  B {\bf 580}, 485 (2000)
[hep-ph/9912329].

\bibitem{LIdependent}
R.~N.~Lee,
  JHEP {\bf 0807}, 031 (2008)
  [0804.3008 [hep-ph]].

\bibitem{AIR}
C.~Anastasiou and A.~Lazopoulos,
JHEP {\bf 0407}, 046 (2004)
[hep-ph/0404258].

\bibitem{FIRE}
A.~V.~Smirnov,
JHEP {\bf 0810}, 107 (2008)
[0807.3243 [hep-ph]].

\bibitem{Reduze}
C.~Studerus,
Comput.\ Phys.\ Commun.\  {\bf 181}, 1293 (2010)
[0912.2546 [physics.comp-ph]].

\bibitem{SmirnovPetukhov}
A.~V.~Smirnov and A.~V.~Petukhov,
Lett.\ Math.\ Phys.\  {\bf 97}, 37 (2011)
[arXiv:1004.4199 [hep-th]].

\bibitem{Roiban:2004yf}
R.~Roiban, M.~Spradlin and A.~Volovich,
Phys.\ Rev.\ D70:026009 (2004)
[hep-th/0403190].

\bibitem{Vergu:2006np}
C.~Vergu,
Phys.\ Rev.\ D75:025028 (2007)
[hep-th/0612250].

\bibitem{MasonSkinner}
L.~J.~Mason and D.~Skinner,
JHEP {\bf 1001}, 064 (2010)
[0903.2083 [hep-th]];\\
JHEP {\bf 1012}, 018 (2010)
[1009.2225 [hep-th]].


\end{thebibliography}
\end{document}